\documentclass{aa}  
 \usepackage[varg]{txfonts}

\usepackage{orcidlink}

\usepackage{natbib}
\bibpunct{(}{)}{;}{a}{}{,}
\usepackage{graphicx}
\usepackage{txfonts}
\usepackage{url}
\usepackage{hyperref}
\hypersetup{
     colorlinks=true, 
     linkcolor=red,  
     filecolor=magenta,   
     citecolor=blue,   
     urlcolor=blue
     }
\begin{document}

   \title{Hubble Space Telescope proper motions of Large Magellanic Cloud star clusters}

   \subtitle{I. Catalogues and results for NGC\,1850 \thanks{Based on proprietary and archival observations with the NASA/ESA \textit{Hubble} Space Telescope, obtained at the Space Telescope Science Institute, which is operated by AURA, Inc., under NASA contract NAS 5-26555.}}

      \author{F. Niederhofer\inst{1}\orcidlink{0000-0002-4341-9819}
          \and
          A. Bellini\inst{2}\orcidlink{0000-0003-3858-637X}
          \and
          V. Kozhurina-Platais\inst{2}\fnmsep\inst{3}\orcidlink{0000-0003-0218-386X}
          \and 
          M. Libralato\inst{4}\fnmsep\inst{5}\orcidlink{0000-0001-9673-7397}
          \and
         M. Häberle\inst{6}\orcidlink{0000-0002-5844-4443}
         \and
         N. Kacharov\inst{1}\orcidlink{0000-0002-6072-6669}
         \and
         S. Kamann\inst{7}\orcidlink{0000-0001-6604-0505}
         \and 
         N. Bastian\inst{8}\fnmsep\inst{9}\orcidlink{0000-0001-5679-4215}
         \and
         I. Cabrera-Ziri\inst{10}\orcidlink{0000-0001-9478-5731}
         \and
         M.-R. L. Cioni\inst{1}\orcidlink{0000-0002-6797-696X}
         \and 
         F. Dresbach\inst{11}\orcidlink{0000-0003-0808-8038}
         \and
         S. Martocchia\inst{10}\orcidlink{0000-0001-7110-6775}
         \and
         D. Massari\inst{12}\orcidlink{0000-0001-8892-4301}
         \and
         S. Saracino\inst{7}\orcidlink{0000-0003-4746-6003}
      }

   \institute{Leibniz-Institut für Astrophysik Potsdam, An der Sternwarte 16, D-14482 Potsdam, Germany\\
        \email{fniederhofer@aip.de}
        \and Space Telescope Science Institute, 3700 San Martin Drive, Baltimore, MD 21218, USA\
        \and Eureka Scientific, Inc., 2452 Delmer Street Suite 100, Oakland, CA 94602-3017, USA\
        \and AURA for the European Space Agency (ESA), Space Telescope Science Institute, 3700 San Martin Drive, Baltimore, MD 21218, USA\
        \and INAF - Osservatorio Astronomico di Padova, Vicolo dell'Osservatorio 5, Padova I-35122, Italy\
        \and Max-Planck-Institut für Astronomie, Königstuhl 17, D-69117 Heidelberg, Germany\
        \and Astrophysics Research Institute, Liverpool John Moores University, IC2 Liverpool Science Park, 146 Brownlow Hill, Liverpool L3 5RF, UK\
        \and Donostia International Physics Center (DIPC), Paseo Manuel de Lardizabal, 4, 20018, Donostia-San Sebastián, Guipuzkoa, Spain\
        \and IKERBASQUE, Basque Foundation for Science, 48013, Bilbao, Spain\
        \and Astronomisches Rechen-Institut, Zentrum für Astronomie der Universität Heidelberg, Mönchhofstraße 12-14, D-69120 Heidelberg, Germany
        \and Lennard-Jones Laboratories, School of Chemical and Physical Sciences, Keele University, Keele ST5 5BG, UK\
       \and INAF – Osservatorio di Astrofisica e Scienza dello Spazio di Bologna, Via Gobetti 93/3, 40129 Bologna, Italy
}

   \date{Received date /
Accepted date}

 
  \abstract{We present proper motion (PM) measurements for a sample of 23 massive star clusters within the Large Magellanic Cloud using multi-epoch data from the \textit{Hubble} Space Telescope (HST). We combined archival data from the ACS/WFC and WFC3/UVIS instruments with observations from a dedicated HST programme, resulting in time baselines between 4.7 and 18.2~yr available for PM determinations. For bright well-measured stars, we achieved nominal PM precisions of 55~$\mu$as\,yr$^{-1}$ down to 11~$\mu$as\,yr$^{-1}$. To demonstrate the potential and limitations of our PM data set, we analysed the cluster NGC~1850 and showcase a selection of different science applications. The precision of the PM measurements allows us to disentangle the kinematics of the various stellar populations that are present in the HST field. The cluster has a centre-of-mass motion that is different from the surrounding old field stars and also differs from the mean motion of a close-by group of very young stars. We determined the velocity dispersion of field stars to be $0.128\pm0.003$~mas\,yr$^{-1}$ (corresponding to $30.3\pm0.7$~km\,s$^{-1}$). The velocity dispersion of the cluster inferred from the PM data set most probably overestimates the true value, suggesting that the precision of the measurements at this stage is not sufficient for a reliable analysis of the internal kinematics of extra-galactic star clusters. 
  Finally, we exploit the PM-cleaned catalogue of likely cluster members to determine any radial segregation between fast and slowly-rotating stars, finding that the former are more centrally concentrated. 
  With this paper, we also release the astro-photometric catalogues for each cluster.}


   \maketitle
%

\section{Introduction}\label{sec:intro}

Within the last years, stellar proper motion (PM) measurements have become a powerful tool in modern astrophysics. Of special interest are globular clusters (GCs), since their internal kinematics as well as their dynamics within their host galaxy provide a treasure chest for studies in the fields of star cluster formation and galaxy evolution. The availability of precise PMs over the entire sky provided by the \textit{Gaia} mission \citep{Gaia16} has given this topic an enormous boost. Combining the on-sky motions of the clusters with spectroscopic line-of-sight velocities and distance information has allowed to trace the orbits of GCs within our Galaxy \citep[e.g.][]{Gaia18, Baumgardt19}. Based on the kinematics, paired with information about the ages and chemical compositions of the clusters, it has been shown that the GCs are grouped in different sub-populations, which enabled to portray the assembly history of our Galaxy \citep[e.g.,][]{Massari19, Kruijssen20}.

The internal dynamics of GCs contain vital information for the understanding of the formation and evolution of these objects. PM measurements from the \textit{Gaia} mission, as well as from the \textit{Hubble} Space Telescope (HST) thereby have proven to be a key ingredient in this respect. Using multi-epoch HST data sets, \citet{Bellini14} and, more recently \citet{Libralato22}, compiled comprehensive catalogues of high-precision internal PMs for a sample of 22 and 56 Galactic GCs, respectively. 
The wealth of astrometric data from \textit{Gaia} and HST has been used for investigations of a variety of kinematic properties, including on-sky rotation of GCs \citep[e.g.][]{AndersonKing03, Massari13, Bellini17b}, internal velocity dispersion profiles \citep[e.g.][]{Watkins15a, Watkins15b, Libralato18, Raso20, Haberle21, Vasiliev21} and velocity anisotropy profiles \citep[e.g.][]{Libralato22}. 

An inherent property of GCs is the presence of multiple stellar populations 
in the form of star-to-star abundance variations in light elements \citep[see][]{Bastian18, Gratton19} and sometimes even in He \citep[see, e.g.][]{Bedin04, Piotto07} and Fe \citep[see, e.g.][]{Marino09, Yong14}. These variations are discrete, resulting in distinct sequences in colour-magnitude diagrams (CMDs) based on appropriate filters. Despite large effort in explaining this phenomenon, its origin is still a mystery. Investigations of the internal kinematics of the different populations have the potential to shed light on the origin of multiple populations and, in turn, further our knowledge of the formation and evolution of GCs. A number of studies have been dedicated to this topic \citep[e.g.][]{Bellini15, Bellini18, Milone18b, Libralato19, Libralato23c} drawing the consistent picture that the stars of the enriched, or second population, become radially anisotropic in the outer regions of the clusters. This kinematic signature suggests that the second population stars were initially more centrally concentrated than first population stars.

Stellar PM measurements also had a significant impact on our understanding of our two neighbouring galaxies, the Large and Small Magellanic Clouds (LMC and SMC). Both galaxies are currently interacting with each other and with the Milky Way. The SMC is thereby in the early phases of a minor-merger event with the LMC. Using multi-epoch HST data, \citet{Kallivayalil06a, Kallivayalil06b, Kallivayalil13} measured for the first time the motion of the Magellanic Clouds in the plane of the sky, showing that they move considerably faster than expected. This fast motion implies that both galaxies have not been long-term companions of the Milky Way, but rather are on their first or second passage around our Galaxy \citep[e.g.][]{Patel17, Vasiliev23}. The internal PM fields of the galaxies are characterised by signs of their past mutual interactions \citep[see, e.g.][]{Zivick18, Zivick19, Niederhofer18b, Niederhofer21, Schmidt20, Choi21}.

The LMC and SMC also harbour a rich population of massive star clusters spanning the full cosmic age range. Studying their kinematics will provide us with valuable insights into the formation history of the Magellanic Clouds as well as the formation of their system of star clusters. However, investigations of the dynamics of star clusters in the Magellanic Clouds using PMs are still in their infancy. Such studies have mostly been hampered by the PM precision required to discriminate the motion of cluster stars from field stars (typically $<$0.1~mas\,yr$^{-1}$ at the distance of the LMC) and the small angular extent of the clusters and the associated high stellar density (owing to the large distances to the objects). 
In a first approach to investigate the kinematic structure of the old star-cluster system of the LMC, \citet{Piatti19} analysed literature line-of-sight velocities and PM measurements from the second data release (DR2) from \textit{Gaia} \citep{Gaia18b}. They found two distinct kinematic groups with different spatial distributions, at odds with what has been found using solely line-of-sight velocities \citep[e.g.][]{Freeman83, Grocholski06, Sharma10}.
Later, \citet{Bennet22} studied the kinematics of the LMC star-cluster population, using a combination of \textit{Gaia} and HST data to derive the bulk PMs of the clusters. They found that the clusters follow disc-like kinematics with no evidence of a halo population. Exploiting multi-epoch HST data sets with long temporal baselines, \citet{Massari21} measured stellar PMs in a field containing the SMC star cluster NGC~419. Thanks to the high precision of their measurements, the authors were able to resolve the kinematics of the different stellar populations within the field \citep[see also][]{Sabbi22, Milone23a}. \citet{Dresbach2022, Dresbach2023} used the astrometric data from \citet{Massari21} to get
clean samples of cluster members of NGC~419 to investigate various features in its CMD. In the first-mentioned study, the authors analysed the population of blue straggler stars in the cluster, finding that the cluster is still in a dynamically young state. In the latter work, they showed that stars across the extended main-sequence turnoff follow the same radial distribution, suggesting that this feature is caused by stars rotating at different velocities rather than a spread in age.

\begin{figure}[t]
\centering
\includegraphics[width=\columnwidth]{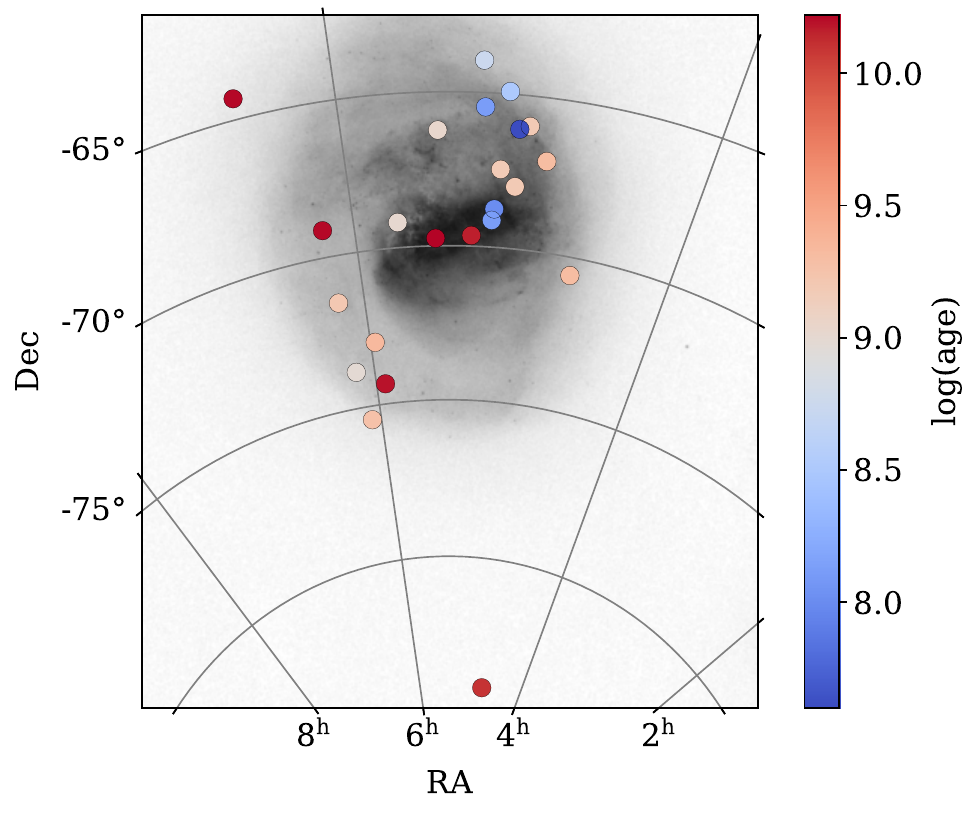}
\caption{Spatial distribution of the 23 LMC star clusters in our sample, colour-coded by their age. The background grey-scale image shows a stellar density map from \textit{Gaia} DR3.\label{fig:cluster_map}}
\end{figure}

Encouraged by the above studies and the scientific prospects, we started an observational campaign using multi-epoch HST data to measure PMs towards a sample of star clusters within the LMC. 
The targeted clusters cover an age range between 40~Myr and 13~Gyr and have distances up to 13~kpc from the centre of the LMC (see Fig.~\ref{fig:cluster_map} and Table~\ref{tab:cluster_param}). In this paper we present the compilation of the PM catalogues and showcase the range of possibilities of the PM data set by analysing the young cluster NGC~1850.

The paper is organised as follows. In Section~\ref{sec:data}, we describe the compilation of the data sets as well as the photometric and astrometric reduction. In Section~\ref{sec:cluster_profiles}, we present the determination of the cluster centres and structural profiles. We describe the methods to measure, correct and calibrate the PMs in Section~\ref{sec:PM}. In Section~\ref{sec:NGC1850_demo} we demonstrate a number of different science applications of our PM data set. In Section \ref{sec:conclusions}, we provide a summary of the paper and offer future prospects.


\section{Data sets and reduction\label{sec:data}}

\subsection{Data sets}

The data from our dedicated HST programme GO-16748 (PI: Niederhofer) add an additional epoch of HST observations for a total of 19 massive (${\rm log}(M/M_{\sun})\gtrsim4.3$) star clusters in the LMC. These clusters either had only a singe epoch of existing data or the time baseline between multiple epochs of archival observations was too short for PM determinations. Since our aim is to determine precise PMs for as many stars as possible, our programme was specifically tailored for astrometric studies. 
The archival data sets are usually customised to specific science cases and have different quality requirements that mainly focus on photometric precision and not necessarily on astrometry. Thus, with our observational setup we ensure to have at least one epoch of observations that is best-suited for astrometric measurements.
In particular, each cluster was observed for one HST orbit using the Ultraviolet-Visible (UVIS) channel of the Wide Field Camera 3 (WFC3) in the F814W filter, using a sequence of properly dithered long and short exposures. We combine these observations with already existing archival data taken with the Wide-Field Channel (WFC) of the Advanced Camera for Surveys (ACS) and WFC3/UVIS. 
In addition to the 19 clusters included in our programme, we present here PMs for four more LMC clusters with suitable archival long time-baseline data.

We note here that not all available data sets are equally usable for PM measurements and, as a consequence, observations taken with specific filters and instruments are not included in our PM determination. As discussed in \citet{Bellini18}, filters bluer than F336W show strong chromatic effects that lead to significant colour dependencies in the distortion solution \citep[see also figure 6 of][and the associated discussion]{Bellini11}. Thus observations in the F275W filter are not suited for high-precision astrometry and we discard these images from our PM calculations.
Also, exposures in narrow-band filters, such as F343N or F656N, are not considered, given the lack of high-precision distortion solutions and their shallow image depth. We further exclude observations taken with the Wide-Field Planetary Camera 2 (WFPC2) due to the much lower image quality compared to ACS or WFC3.
Our final assembled set of multi-epoch observations provides temporal baselines between 4.7 and 18.2~yr, sufficiently long for precise PM measurements. 
A full detailed journal of all observations used to compute the PMs for the 23 clusters presented in this study is given in Appendix~\ref{app:obs}. 
Additionally, all the HST data used in this paper can be found in the Mikulski Archive for Space Telescopes (MAST) under the following DOI: \href{https://doi.org/10.17909/7d5e-s940}{10.17909/7d5e-s940}.

The photometric reduction process closely follows the methods presented in \citet{Bellini17, Bellini18, Nardiello18} and \citet{Libralato22}. We provide here a brief outline of the main steps and direct the interested reader to the aforementioned papers for a detailed description.

\subsection{First-pass photometry}

Our analysis is based on \texttt{$\_$flc} images that contain the un-resampled pixel data. The images are processed through the standard HST calibration pipeline and are dark- and bias-corrected and flat-fielded. Also pixel-based imperfect charge transfer efficiency (CTE) corrections have been applied to the exposures, as described in \citet{Anderson18} and \citet{Anderson21}.

In a first step, we used the \texttt{Fortran} routine \texttt{hst1pass} \citep{Anderson22} to create an initial catalogue of bright objects. The code performs a single run of source finding and PSF fitting without neighbour subtraction on each exposure individually. To account for time-dependent variations of the PSFs that are caused by telescope breathing effects, we perturbed the spatially variable library PSFs. We started with the ``PSF By Focus'' (PBF) models\footnote{\href{https://www.stsci.edu/~jayander/HST1PASS/LIB/PSFs/STDPBFs/}{https://www.stsci.edu/~jayander/HST1PASS/LIB/PSFs/STDPBFs/}} to first determine the focus level of the telescope. Then we used the PSF models at the appropriate focus level to derive spatially variable perturbations to the PSFs using a sample of bright, unsaturated and isolated stars. The perturbation models were derived on grids that range from 1~$\times$~1 to 5~$\times$~5 grid cells, depending on the number of stars available for the fitting of the perturbations. The final PSF models that are now tailored specifically to each exposure were then used to measure the positions and fluxes of the stars in each image. Saturated stars were measured with \texttt{hst1pass} using the methods described in \citet{Gilliland04} and \citet{Gilliland10}.
Finally, stellar positions were corrected for geometric distortion applying the distortion solutions provided by \cite{AndersonKing06} for ACS/WFC\footnote{We also used a look-up table of residuals to account for changes in the original distortion solution of the ACS/WFC that occurred during the \textit{Hubble} Service Mission 4 in 2009.}, and \citet{Bellini09} and \citet{Bellini11} for WFC3/UVIS. Several clusters in our sample have observations in the WFC/UVIS filter F475W, for which no high-precision distortion correction exists. To also incorporate these data in our PM determinations, we derived our own distortion solution for this filter.

\subsection{Master frame}

In the next step, we define a distortion-free common frame of reference to which we relate the photometry and astrometry of all catalogues from the first-pass photometric run. This frame is referred to as the master frame. We set up the master frame such that the X and Y axes are aligned with the west and north directions, respectively. The pixel scale is chosen to be 40~mas\,pixel$^{-1}$, close to that of WFC3/UVIS (which is 39.77~mas\,pixel$^{-1}$, see \citealt{Bellini09, Bellini11}). Stellar positions are registered to the $Gaia$ data release 3 (DR3) astrometric reference frame \citep{Gaia23}. For this, we first shifted the positions of the \textit{Gaia} sources to the locations at the mean observation date of each HST epoch using the PMs in the $Gaia$ catalogue (the positions as given in the DR3 catalogue refer to epoch 2016.0). Then, we used a tangent-plane projection to de-project the $Gaia$ coordinates to a flat plane using the centres of the clusters\footnote{We used the centre coordinates from the Simbad data base: \href{https://simbad.u-strasbg.fr/simbad/}{https://simbad.u-strasbg.fr/simbad/}} as the tangent point. To avoid negative coordinates in the master frame, we further shifted the positions of the cluster centres to pixel coordinates (10\,000, 10\,000). 

We created master-frame catalogues individually for each filter and epoch. To do so, we first cross-identified bright, unsaturated stars from our single-exposure catalogues with the \textit{Gaia} DR3 catalogue to determine initial estimates of general six parameter transformations to the master frame. We applied these transformations to all stars in the catalogues. Then, we improved the transformations by averaging the master frame positions within the longest exposures in each filter and using these positions as the new master-frame coordinates. We iterated a few times, using only bright, well-measured and unsaturated stars as reference objects, until the transformation residuals did not improve any more. To create the final first-pass catalogue for each filter, we averaged together the master-frame positions of the stars coming from the individual exposures. The instrumental magnitudes of the stars from the individual exposures were first zero-pointed to the longest available exposure in that filter before averaging them together.

\subsection{Second-pass photometry}

We then performed a second-pass photometric run using the \texttt{Fortran} software \texttt{KS2} developed by J. Anderson (see \citealt{Sabbi16}, \citealt{Bellini17} and \citealt{Nardiello18} for a detailed description of the functionality of the code). \texttt{KS2} uses the products from the first-pass photometry, like the PSF models and the transformations of each exposure to the master frame, to simultaneously find and measure stars in all individual exposures coming from all filters. It is also able to analyse data coming from multiple HST instruments at once. \texttt{KS2} employs multiple iterations to find and measure stars. Starting from the brightest sources, \texttt{KS2} iteratively identifies sources, determines their positions and fluxes and subtracts them from the images. Within each iteration, the routine searches for fainter stars that have not been detected in the previous iteration. With the addition of the newly found stars, also the fitting of the already found stars is improved within each iteration, by including the profiles of neighbouring stars in the fit. \texttt{KS2} further uses the catalogues resulting from the first-pass photometry to create masks around bright and saturated stars to lower the detection of spurious sources in the diffraction spikes and PSF features of these objects. 

As discussed in \citet{Bellini18}, \texttt{KS2} measures sources using three different methods. However, only in method 1, both, stellar fluxes and positions are measured in each (neighbour-subtracted) exposure fitting the local PSF model. Thus, this method is best suited for astrometric studies and is used in this work for our PM measurements. 

For each cluster, we grouped together all observations that are taken roughly within the same year and ran \texttt{KS2} separately on these different epochs, such that we ended up with one master catalogue for each epoch. 
Within \texttt{KS2}, the user can specify the number of finding and measuring iterations and also which set of exposures will be used for the source finding within a given iteration. Since observations for each cluster are composed of very heterogeneous data sets,
with a broad range of used filters and exposures, we opted for individual iteration strategies that are best-suited for each set of epoch and cluster.  

Besides the fluxes and positions, \texttt{KS2} also provides for each source a number of different quality parameters that can be employed to select a sub-sample of well-measured stars. There are:
\begin{itemize}
    \item the photometric error, defined as the root-mean-square of measurements coming from individual exposures, \texttt{RMS};
    \item the quality-of-fit parameter \texttt{QFIT}, which indicates how well a source has been fitted by the PSF model. It relates the actual measured pixel values to the expected values from the PSF model. The values of \texttt{QFIT} are between 0 and 1, where a value of 1 corresponds to a perfect fit;
    \item the shape parameter \texttt{RADXS} \citep[see][]{Bedin08}, which is a measure of the amount of the flux of a source outside the PSF core compared to the expected flux from the PSF model. Extended sources, such as galaxies and stellar blends have positive values whereas sources sharper than the PSF, like cosmics and artefacts, have negative values;
    \item the isolation parameter $o$ which reports the fraction of flux, before neighbour subtraction, within the PSF fitting aperture, which is due to neighbouring sources;
    \item the number of exposures a star is detected, N$_{\rm f}$, and the final number of exposures that are actually used to measure its position and flux, N$_{\rm g}$.
\end{itemize}

\subsection{Photometric calibration}

Following the procedures described in \citet{Bellini17} and \citet{Nardiello18}, we calibrate the photometry for each filter and epoch to the VEGA-mag system by comparing the output magnitudes from \texttt{KS2} to magnitudes resulting from aperture photometry on \texttt{\_drc} images. For a given filter, the calibrated magnitude $m_{\rm Cal}$ is calculated as:
\begin{equation}
    m_{\rm Cal} = m_{\rm Inst} + \Delta m + \rm ZP,
\end{equation}
\noindent
where $m_{\rm Inst}$ is the instrumental magnitude as measured by \texttt{KS2}, $\Delta m$ is the 2.5$\sigma$-clipped median difference between the magnitudes resulting from aperture photometry, $m_{\rm Ap}$, and $m_{\rm Inst}$, and ZP is the photometric zero-point in the VEGA-mag system.

We performed aperture photometry on the \texttt{$\_$drc} images (which are normalised to an exposure time of 1\,sec) and considered only bright, unsaturated and isolated stars (no brighter neighbour within 20 pixels) for the calibration. We measured the flux within an aperture of 10 pixels (0\farcs4 for WFC3/UVIS and 0\farcs5 for ACS/WFC) and determined the local sky level within an annulus between 12 and 16 pixels. To correct the photometry for the finite aperture, we used the corresponding encircled energy fractions as given on the STScI instrument webpages\footnote{\href{https://www.stsci.edu/hst/instrumentation/acs/data-analysis/aperture-corrections}{https://www.stsci.edu/hst/instrumentation/acs/data-analysis/aperture-corrections} for ACS/WFC and \href{https://www.stsci.edu/hst/instrumentation/wfc3/data-analysis/photometric-calibration/uvis-encircled-energy}{https://www.stsci.edu/hst/instrumentation/wfc3/data-analysis/photometric-calibration/uvis-encircled-energy} for WFC3/UVIS}.
For each filter and epoch, we determined the quantity $\Delta m$ by cross-matching the aperture-photometric catalogue with the PSF-based catalogue from \texttt{KS2} and calculating the 2.5$\sigma$-clipped median differences of the magnitudes of the stars in common. For observations taken with ACS/WFC, we obtained the zero-points using the \texttt{python} implementation of the ``ACS Zeropoints Calculator''\footnote{\href{https://www.stsci.edu/hst/instrumentation/acs/data-analysis/zeropoints}{https://www.stsci.edu/hst/instrumentation/acs/data-analysis/zeropoints}}. For WFC3/UVIS, we determined the time-dependent zero-points following the procedure described by \citet{Calamida22}.


\section{Cluster centres and structural profiles \label{sec:cluster_profiles}}

\begin{figure}[t]
\centering
\includegraphics[width=\columnwidth]{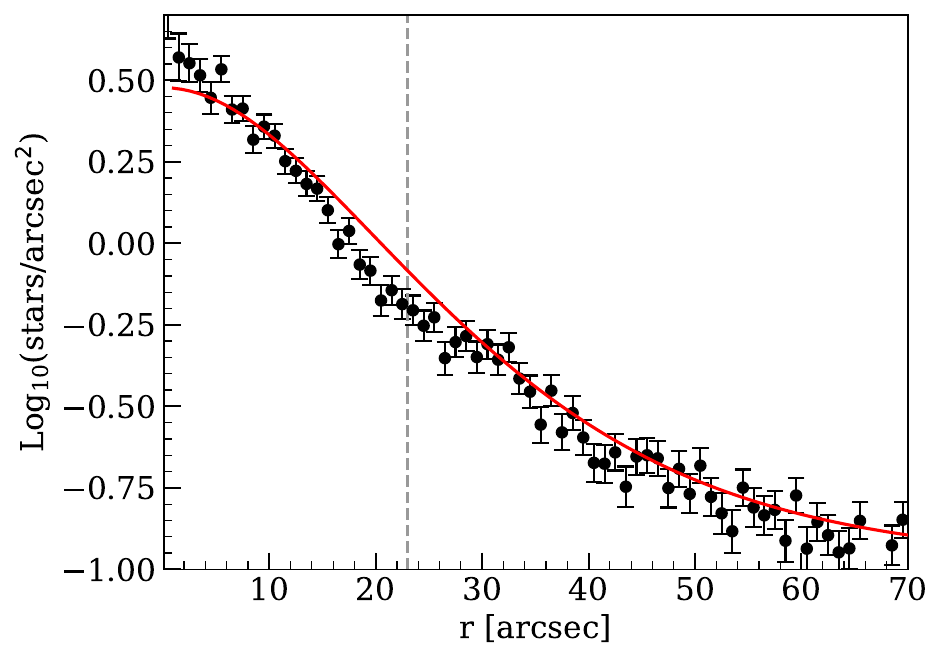}
\caption{Radial surface density profile of NGC~1850 created from elliptical annuli of size 1~arcsec (individual points with error bars). The red solid line follows the best-fit Plummer profile. The vertical dashed line indicates the location of the scale radius $r_h$.\label{fig:plummer_plot}}
\end{figure}

For the clusters in our sample we also derive the clusters' centres and structural parameters, using the discrete maximum likelihood approach as described in detail in \citet{Martin08} and \citet{Kacharov14}\footnote{\href{https://github.com/kacharov/morphology_2d}{https://github.com/kacharov/morphology\_2d}}. To describe the stellar density profile of the clusters, we used a Plummer profile \citep{Plummer1911} of the following form: 
\begin{equation}
    n(r)=n_0\left( 1+\frac{r^2}{r_h^2}\right)^{-2} + n_f, 
\end{equation}
where $n_0$ is the central number density of the cluster, $r_h$ is the scale radius of the cluster (in this case the radius within which half of the cluster stars are), and $n_f$ is the field stellar density. We allow for an elliptical profile, thus the radius $r$ is considered as an elliptical radius. For the clusters included in our GO-16748 programme, we used the \texttt{KS2} master frame catalogues from this epoch for the fitting. For the remaining four clusters, we used the deep observations from programme GO-10595 (NGC~1806 and NGC~1846), GO-15630 (NGC~1978) and GO-16255 (NGC~1783). We note here that our final astro-photometric catalogues are not suited for determining the structural parameters, since these catalogues only contain stars for which PMs have been calculated and thus do not necessarily provide a homogeneous level of completeness across the total field-of-view, even at bright magnitudes. For all clusters, we selected all stars brighter than 1 magnitude below the main sequence turn-off for the fitting. We simultaneously fit for six free parameters: the centroid coordinates of the cluster ($x_0$, $y_0$), its scale radius ($r_h$), the ellipticity ($\epsilon$), the position angle of the major axis ($\theta$, measured from north to east), and the field star density ($n_f$). We explored the posterior probability space using the Markov Chain Monte Carlo (MCMC) sampler \texttt{emcee} \citep{Foreman-Mackey13} which is a \texttt{python} implementation of the affine-invariant MCMC ensemble sampler \citep{Goodman10}. We ran the MCMC for 600 steps using an ensemble of 40 walkers and adopted the mean and the standard deviations of the last 25 per cent of the chain to be the best-fitting values of the free parameters and their uncertainties, respectively. As an example, Fig.~\ref{fig:plummer_plot} shows the radial surface density profile of NGC~1850 together with the best-fit Plummer profile. We finally transformed the pixel coordinates of the centroid to RA and Dec coordinates using the derived WCS for each cluster (see Section~\ref{subsec:astrometry}). The results for all clusters are presented in Table~\ref{tab:cluster_param}.

\begin{table*}\small
\centering
\caption{Parameters of the LMC star clusters. \label{tab:cluster_param}}
\begin{tabular}{lccccccccccccr}
\hline\hline
\noalign{\smallskip}
Cluster ID & RA$_0$ & $\Delta$RA$_0$ & Dec$_0$ & $\Delta$Dec$_0$ & Age & (m--M)$_0$ & D$_{\rm LMC}$ & r$_{\rm h}$ & $\Delta$r$_{\rm h}$ & $\epsilon$ & $\Delta\epsilon$ & $\theta$ & $\Delta\theta$\\
& [h:m:s] & [arcsec] & [\degr:\arcmin:\arcsec] & [arcsec] & [Gyr] & [mag] & [kpc] & [arcsec] & [arcsec] & & & [\degr] & [\degr] \\
\noalign{\smallskip}
\hline
\noalign{\smallskip}
Hodge 11 & 06:14:22.83 & 0.38 & $-$69 50 50.0 & 0.38 & 13.4 & 18.57 & 4.63 & 29.71 & 0.66 & 0.04 & 0.03 & 25 & 16 \\
NGC 1651 & 04:37:32.29 & 0.46 & $-$70 35 10.5 & 0.43 & 2.05 & 18.48 & 3.34 & 26.82 & 1.04 & 0.01 & 0.04 & -- & -- \\
NGC 1718 & 04:52:26.02 & 0.37 & $-$67 03 05.1 & 0.31 & 1.85 & 18.43 & 3.22 & 28.73 & 0.70 & 0.04 & 0.02 & 49 & 15 \\
NGC 1783 & 04:59:08.94 & 0.38 & $-$65 59 14.8 & 0.34 & 1.65 & 18.51 & 3.40 & 43.77 & 0.81 & 0.10 & 0.01 & 1  & 1  \\
NGC 1805 & 05:02:21.66 & 0.34 & $-$66 06 42.4 & 0.31 & 0.045& 18.32 & 4.82 & 11.00 & 0.50 & 0.01 & 0.05 & -- & -- \\
NGC 1806 & 05:02:12.18 & 0.31 & $-$67 59 10.0 & 0.43 & 1.60 & 18.52 & 1.94 & 28.93 & 0.57 & 0.10 & 0.10 & 28 & 6  \\
NGC 1831 & 05:06:16.19 & 0.38 & $-$64 55 08.8 & 0.43 & 0.70 & 18.41 & 4.31 & 32.45 & 0.67 & 0.16 & 0.02 & 72 & 3  \\
NGC 1841 & 04:45:22.60 & 0.30 & $-$83 59 54.6 & 0.29 & 12.4 & 18.34 &12.85 & 45.21 & 0.46 & 0.11 & 0.01 & 46 & 3  \\
NGC 1846 & 05:07:33.90 & 0.47 & $-$67 27 43.2 & 0.51 & 1.60 & 18.52 & 2.01 & 44.52 & 0.65 & 0.07 & 0.02 & 2  & 2  \\
NGC 1850 & 05:08:45.33 & 0.66 & $-$68 45 40.5 & 0.38 & 0.07 & 18.38 & 2.65 & 22.98 & 0.74 & 0.07 & 0.03 & 68 & 17 \\
NGC 1856 & 05:09:30.17 & 0.28 & $-$69 07 43.6 & 0.24 & 0.25 & 18.32 & 3.85 & 24.38 & 0.63 & 0.04 & 0.03 & 74 & 14 \\
NGC 1866 & 05:13:38.63 & 0.38 & $-$65 27 52.7 & 0.42 & 0.20 & 18.30 & 5.30 & 29.24 & 0.52 & 0.06 & 0.02 & 55 & 10 \\
NGC 1868 & 05:14:36.02 & 0.28 & $-$63 57 15.0 & 0.25 & 1.15 & 18.45 & 4.73 & 15.19 & 0.48 & 0.11 & 0.03 & 12 & 6  \\
NGC 1898 & 05:16:42.04 & 0.18 & $-$69 39 24.2 & 0.17 & 11.7 & 18.60 & 2.61 & 25.75 & 0.30 & 0.07 & 0.01 & 84 & 3  \\
NGC 1978 & 05:28:45.13 & 0.18 & $-$66 14 11.9 & 0.20 & 2.50 & 18.53 & 2.94 & 36.76 & 0.27 & 0.21 & 0.01 & 55 & 1  \\
NGC 2005 & 05:30:10.24 & 0.10 & $-$69 45 10.0 & 0.09 & 13.1 & 18.44 & 1.44 & 12.79 & 0.17 & 0.02 & 0.02 & 61 & 19 \\
NGC 2108 & 05:43:57.00 & 0.48 & $-$69 10 52.0 & 0.46 & 1.00 & 18.48 & 1.88 & 27.85 & 0.99 & 0.18 & 0.03 & 51 & 6  \\
NGC 2173 & 05:57:58.50 & 0.48 & $-$72 58 42.6 & 0.55 & 1.70 & 18.37 & 4.87 & 24.41 & 0.77 & 0.05 & 0.03 & 46 & 2  \\
NGC 2203 & 06:04:42.50 & 0.50 & $-$75 26 15.0 & 0.48 & 1.65 & 18.38 & 6.42 & 33.64 & 0.64 & 0.00 & 0.03 & -- & -- \\
NGC 2209 & 06:08:36.36 & 0.80 & $-$73 50 14.6 & 0.81 & 1.15 & 18.39 & 5.52 & 34.76 & 1.25 & 0.08 & 0.01 & 48 & 10 \\
NGC 2210 & 06:11:31.65 & 0.11 & $-$69 07 18.4 & 0.10 & 12.0 & 18.36 & 4.84 & 16.26 & 0.18 & 0.09 & 0.01 & 9  & 4  \\
NGC 2213 & 06:10:42.19 & 0.35 & $-$71 31 45.6 & 0.36 & 1.60 & 18.36 & 4.99 & 15.92 & 0.52 & 0.04 & 0.04 & -- & -- \\
NGC 2257 & 06:30:12.50 & 0.42 & $-$64 19 37.5 & 0.38 & 11.8 & 18.37 & 7.67 & 41.64 & 0.57 & 0.09 & 0.02 & 42 & 5  \\

\noalign{\smallskip}
\hline

\end{tabular}
\tablefoot{The columns report the centre coordinates (RA$_0$ and Dec$_0$), the age, the distance modulus ((m--M)$_0$), the 3D distance to the centre of the LMC (D$_{\rm LMC}$), the scale radius (r$_{\rm h}$), the ellipticity ($\epsilon$) and the position angle of the major axis ($\theta$, measured from north to east).\\
Values for age and distance modulus are taken from \citet{Milone23a}.\\
The 3D distance from the LMC centre (D$_{\rm LMC}$) is calculated assuming the (m--M)$_0$ values of the clusters as stated in the table, a mean distance to the LMC of 49.9~kpc \citep{deGrijs14} and the coordinates of the dynamical centre of the LMC as determined by \citet{Niederhofer22}.
}
\end{table*}


\section{PM calculations \label{sec:PM}}

\subsection{Relative PMs}

We calculated relative PMs using the sophisticated method that was first developed by \citet{Bellini14} and then later improved by \citet{Bellini18} and \citet{Libralato18, Libralato22}. The process relies on treating each exposure as an independent measurement and the PMs are calculated in an iterative way, with a nested set of ``small'' and ``big'' iterations. We briefly outline the main steps of the procedure here. A detailed description is presented in the above-mentioned works. 

For each cluster, we start by cross-matching the \texttt{KS2} master-frame catalogues at each epoch with the ones from all other epochs to identify those stars that have detections in at least two epochs. This constitutes our master list for the PM calculation. \texttt{KS2} also produces for each single exposure a catalogue of the raw fluxes and positions (including neighbour subtraction) in the detector frame. We used these catalogues for the calculation of the PMs. A big iteration starts by cross-identifying stars in each single-exposure catalogue with the master list. For the first big iteration, a generous searching radius of 2.5 pixels was used. Then, a sequence of small iterations was executed. Within each small iteration, the positions of the stars in the individual exposures were transformed to the master frame by means of a general six-parameter transformation, using bright, unsaturated cluster members as reference stars. In the very first small iteration, the reference stars have been selected based on their positions in the CMD, in subsequent iterations they were selected based on their PMs. Since we used cluster members as reference stars for the transformation, the resulting PMs will be relative to the bulk motion of the cluster. Thus, the distribution of cluster stars in the vector-point diagram (VPD) will be centred at zero. The master-frame transformed X and Y positions of each star were then fitted as a function of the observing time by means of a least-squares straight line. The slope of the fitted line provides a direct measure of the star's PM. This fitting process itself is also iterative and involves rejections of obvious outliers and clipping of data points not in agreement with a Gaussian distribution. The last straight-line fit within a small iteration is performed using locally-transformed stellar master frame positions, based only on the closest 45 reference stars. 
The errors of the PMs are then determined as the uncertainty in the slope of the fitted line, resulting from expected positional uncertainties as a function of the instrumental magnitude \citep[cf. Section 5.2 in][]{Bellini14}. After each small iteration, the sample of reference stars was refined to improve the transformations on the master frame. A big iteration ends when the reference star list does not change any more. At the end of a big iteration, the master list is refined by PM-shifting the positions of its members to positions at a common reference epoch, which, for a given cluster, is chosen to be the average epoch of all its observations. 
For the cross-identification step at the beginning of subsequent big iterations, we PM-shifted the master-frame positions to match the epoch of observation of each individual exposure and also applied a tighter matching radius of 0.5 pixels. This adjustment helps in reducing the number of mismatches. 
The PM determination converges when the change in master-list positions is negligible from one big iteration to the next (typically less than 0.01--0.005 pixels).

We would like to mention here that \texttt{KS2} does not measure saturated stars, which means they are not included in the individual catalogues of those exposures in which they are saturated. They are, however, listed in the master catalogue of a given epoch and filter, even if they are saturated in all corresponding exposures. In this case, the flux and position measurements come from the first-pass photometry. For this reason, our final PM catalogues contain saturated stars as well. The necessary condition for these stars to have a PM measurement is that they are unsaturated in at least four exposures that cover a minimum time baseline of two years.

\begin{figure*}
\begin{tabular}{cc}
\includegraphics[width=\columnwidth]{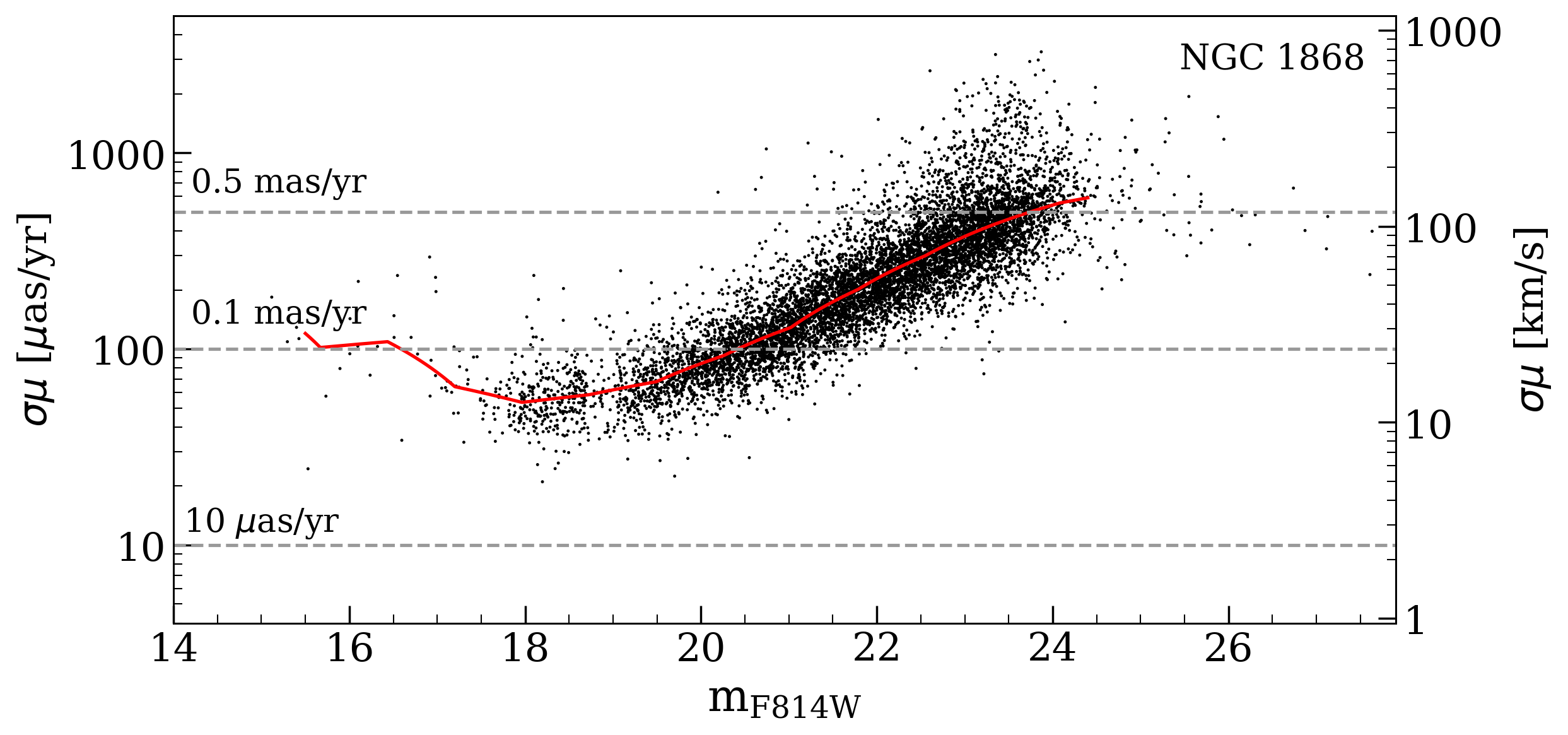} &
\includegraphics[width=\columnwidth]{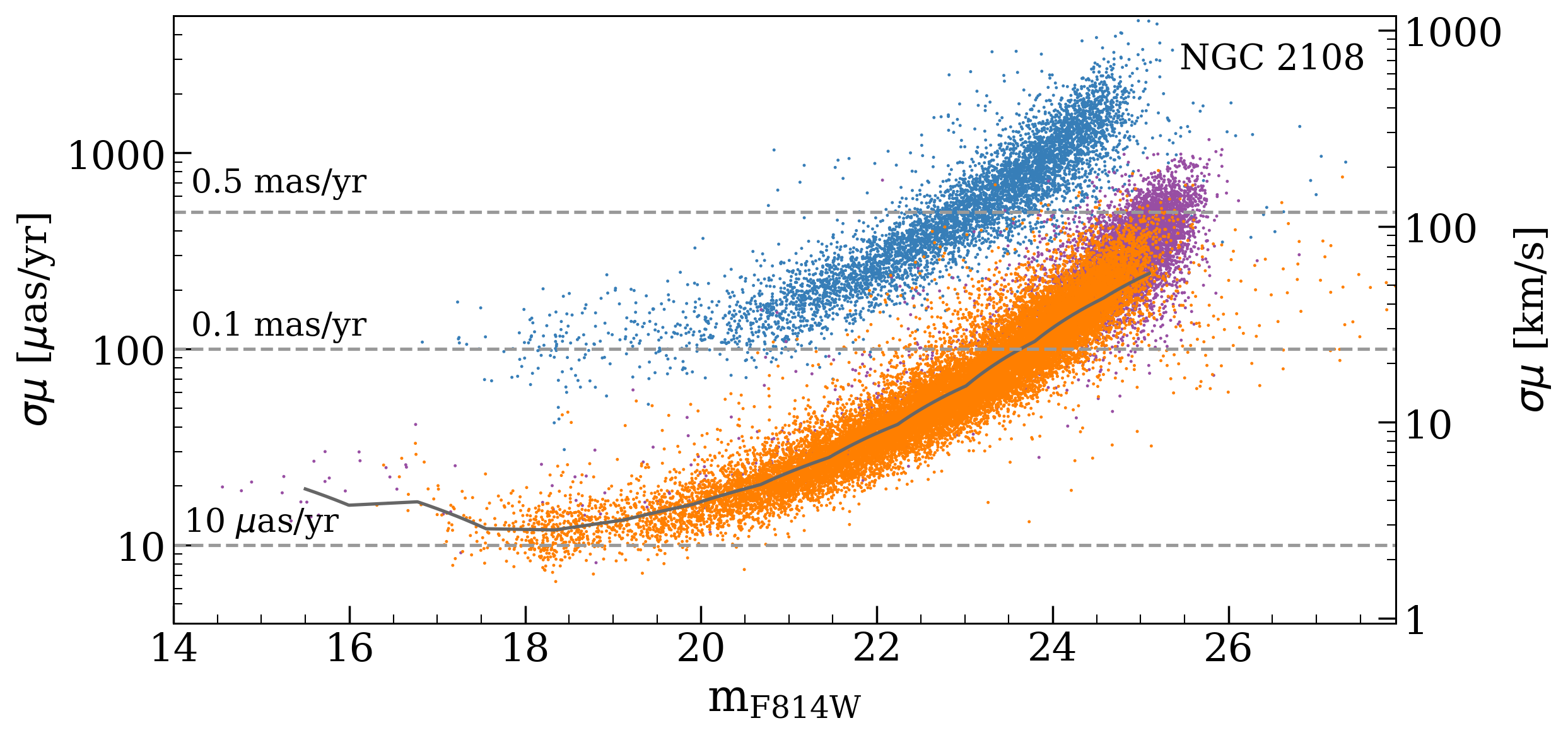} 
\\
\end{tabular}
\caption{Range of PM precisions among the clusters. Shown are the 1D PM uncertainties as a function of the F814W magnitude, for the two clusters NGC~1868 (left panel) and NGC~2108 (right panel). The two clusters have very different data sets available for the PM calculations, in terms of time baseline and numbers of exposures and represent the lower and upper end of the quality scale of our PM measurements. For NGC~1868 we reach a median nominal PM precision for the best-measured stars of about 55~$\mu$as\,yr$^{-1}$, whereas for NGC~2108, the median nominal precision of the best-measured stars is about 11~$\mu$as\,yr$^{-1}$. The solid line in both panels follow the median $\sigma\mu$ as a function of the magnitude. The different sequences in the right-hand panel correspond to PMs determined using time baselines of $\sim$3.0~yr (blue), $\sim$15.1~yr (purple) and $\sim$18.2~yr (orange).
\label{fig:pm_err}
}

\end{figure*}

For the 23 clusters within our sample of star clusters, we determined relative PMs for a total of $\sim$855\,000 stars, with a median of $\sim$32\,000 sources per cluster. The field-of-view of NGC~2209 contains the least number of stars ($\sim$7\,000) whereas the one of NGC~2005 the most ($\sim$112\,000). 
Since the PMs were derived from heterogeneous sets of data, their quality varies among the clusters in our sample. The precision of the PMs is mainly determined by the time baseline and number of exposures used for their calculation. Table~\ref{tab:PM_ZP} shows a record of these values for each cluster. To illustrate the range of the PM quality, Fig.~\ref{fig:pm_err} shows the 1D PM uncertainties as a function of the F814W magnitudes for NGC~1868 and NGC~2108. These two clusters represent the lower and upper end of the quality scale of our PM measurements. The PMs towards NGC~1868 were computed with only 12 individual exposures and a time baseline between the observations of 4.7~yr. For NGC~2108, we have a total of 18 images that are distributed across four epochs, covering a total time baseline of 18.2~yr. For NGC~1868 we reach a median PM precision for the bright well-measured stars of about 55~$\mu$as\,yr$^{-1}$ ($\sim$12.7~km\,s$^{-1}$), whereas for NGC~2108, the median precision of the best-measured stars is about 11~$\mu$as\,yr$^{-1}$ ($\sim$2.8~km\,s$^{-1}$).

\begin{table*}\small
\centering
\caption{Overview of different quantities related to the PM calculations of the LMC clusters. \label{tab:PM_ZP}}
\begin{tabular}{lccccc}
\hline\hline
\noalign{\smallskip}
Cluster ID & $\Delta$time & N$_{\rm epochs}$ & N$_{\rm exp}$ & N$_{\rm sources}$ & ZP($\mu_{\alpha}\rm cos(\delta)$, $\mu_{\delta}$)\\
& [yr] & & & & [mas\,yr$^{-1}$] \\
\noalign{\smallskip}
\hline
\noalign{\smallskip}
Hodge 11 & 5.8  & 2 & 56 & 54\,264& ($-$1.545$\pm$0.041, $-$0.986$\pm$0.046) \\
NGC 1651 & 10.3 & 2 & 13 & 14\,000& ($-$1.956$\pm$0.034, 0.305$\pm$0.038)  \\
NGC 1718 & 9.8  & 2 & 13 & 21\,048& ($-$1.851$\pm$0.040, 0.462$\pm$0.036)  \\
NGC 1783 & 17.3 & 4 & 20 & 40\,944& ($-$1.648$\pm$0.036, 0.025$\pm$0.031)  \\
NGC 1805 & 6.3  & 3 & 13 &  7\,729& ($-$1.593$\pm$0.033, $-$0.100$\pm$0.031) \\
NGC 1806 & 8.2  & 3 & 13 & 33\,892& ($-$1.823$\pm$0.035, 0.040$\pm$0.034)  \\
NGC 1831 & 5.8  & 2 & 14 & 18\,019& ($-$1.714$\pm$0.035, 0.014$\pm$0.033)  \\
NGC 1841 & 6.1  & 3 & 53 & 52\,055& ($-$1.967$\pm$0.032, $-$0.010$\pm$0.035) \\
NGC 1846 & 7.5  & 3 & 19 & 29\,849& ($-$1.745$\pm$0.051, $-$0.176$\pm$0.054) \\
NGC 1850 & 11.9 & 5 & 42 & 63\,024& ($-$2.011$\pm$0.028, $-$0.119$\pm$0.030) \\
NGC 1856 & 8.7  & 3 & 31 & 80\,950& ($-$1.795$\pm$0.035, $-$0.136$\pm$0.035) \\
NGC 1866 & 6.5  & 3 & 29 & 27\,777& ($-$1.558$\pm$0.030, $-$0.177$\pm$0.039) \\
NGC 1868 & 4.7  & 2 & 12 &  8\,694& ($-$1.725$\pm$0.046, 0.038$\pm$0.053)  \\
NGC 1898 & 8.7  & 2 & 14 & 51\,028& ($-$1.962$\pm$0.036, $-$0.274$\pm$0.041) \\
NGC 1978 & 15.9 & 4 & 24 & 59\,762& ($-$1.813$\pm$0.033, $-$0.448$\pm$0.041) \\
NGC 2005 & 10.8 & 2 & 13 & 12\,110& ($-$1.847$\pm$0.038, $-$0.469$\pm$0.043) \\
NGC 2108 & 18.2 & 3 & 18 & 34\,782& ($-$1.610$\pm$0.033, $-$0.728$\pm$0.034) \\
NGC 2173 & 10.7 & 2 & 17 & 12\,642& ($-$1.978$\pm$0.037, $-$0.851$\pm$0.038) \\
NGC 2203 & 11.2 & 2 & 17 & 12\,823& ($-$1.955$\pm$0.033, $-$0.841$\pm$0.038) \\
NGC 2209 & 9.2  & 2 & 12 &  6\,896& ($-$1.852$\pm$0.054, $-$0.936$\pm$0.053) \\
NGC 2210 & 6.1  & 2 & 56 & 63\,257& ($-$1.533$\pm$0.038, $-$1.303$\pm$0.033) \\
NGC 2213 & 10.3 & 2 & 13 &  9\,831& ($-$1.829$\pm$0.046, $-$0.988$\pm$0.042) \\
NGC 2257 & 6.1  & 2 & 54 & 39\,651& ($-$1.467$\pm$0.036, $-$0.970$\pm$0.041) \\

\noalign{\smallskip}
\hline

\end{tabular}

\tablefoot{
The columns report the time baselines ($\Delta$time), number of epochs (N$_{\rm epochs}$) and number of exposures (N$_{\rm exp}$) used to compute the PMs for each cluster. Also given are the number of sources (N$_{\rm sources}$) for which PMs have been determined as well as the PM zero-points needed to transform the relative PMs in our catalogue to an absolute scale.
}

\end{table*}

\subsection{PM corrections\label{subsec:PM_corr}}

As discussed in previous works \citep[e.g.][]{Bellini18, Libralato18, Libralato22}, the raw PMs determined from HST data suffer from systematic effects that vary as a function of position on the master frame. They can be divided in two categories: low- and high-frequency systematic effects. The former is caused by different overlaps of the data sets used to determine the PMs, and thus correlates with varying time baselines and numbers of exposures. The latter effect is due to uncorrected CTE and distortion residuals.

\begin{figure*}
\begin{tabular}{ccc}
\includegraphics[width=0.64\columnwidth]{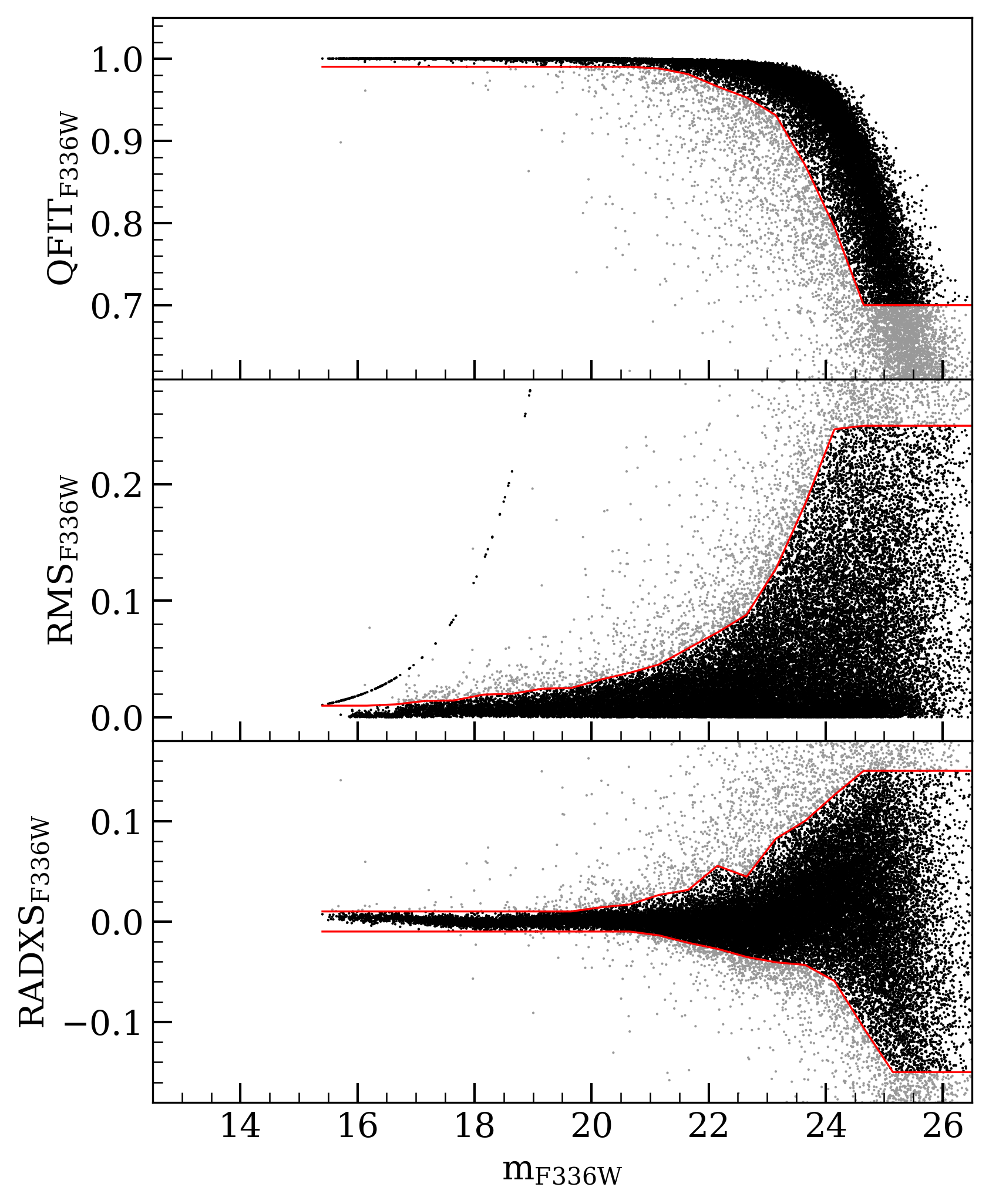} &
\includegraphics[width=0.64\columnwidth]{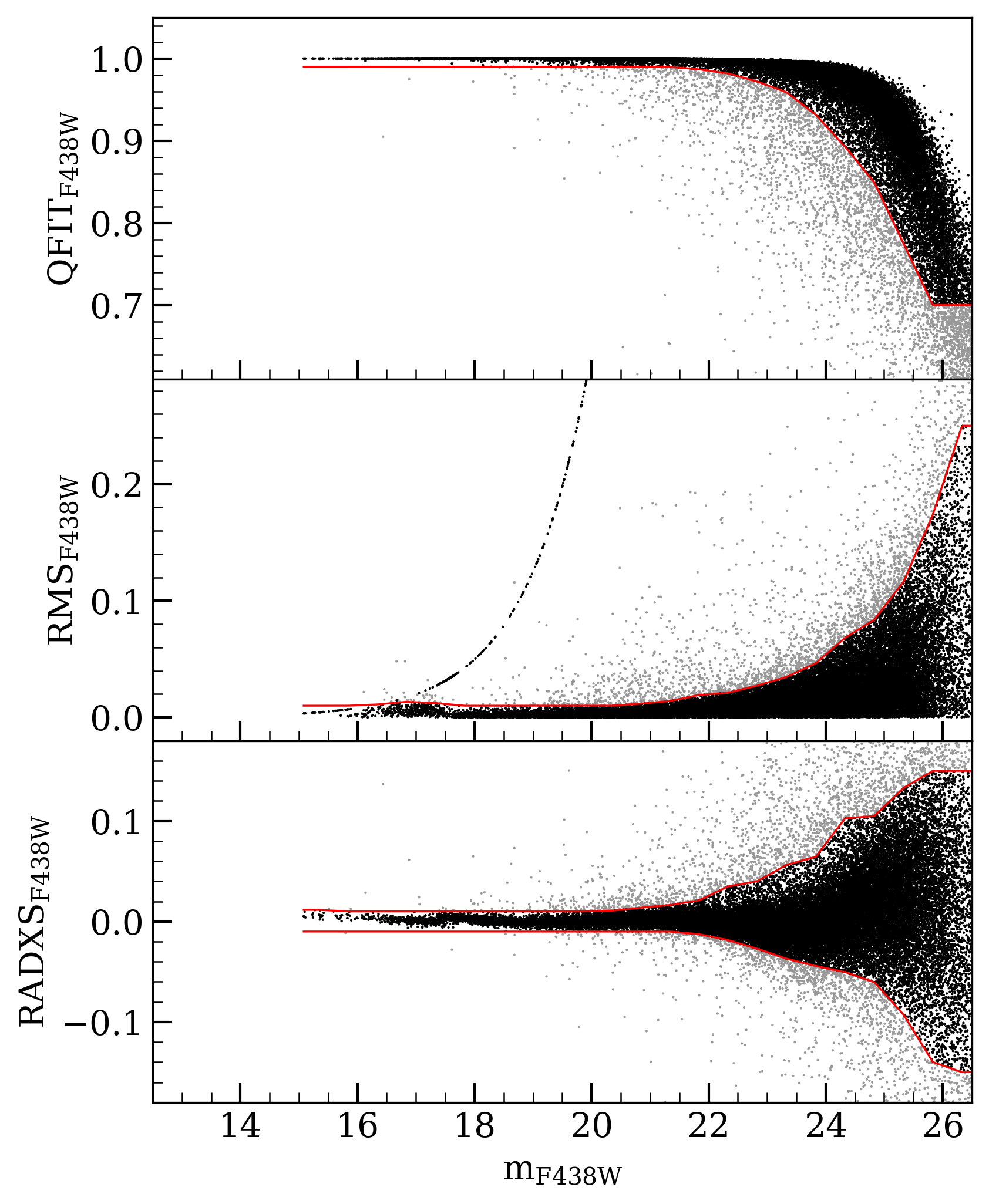} &
\includegraphics[width=0.64\columnwidth]{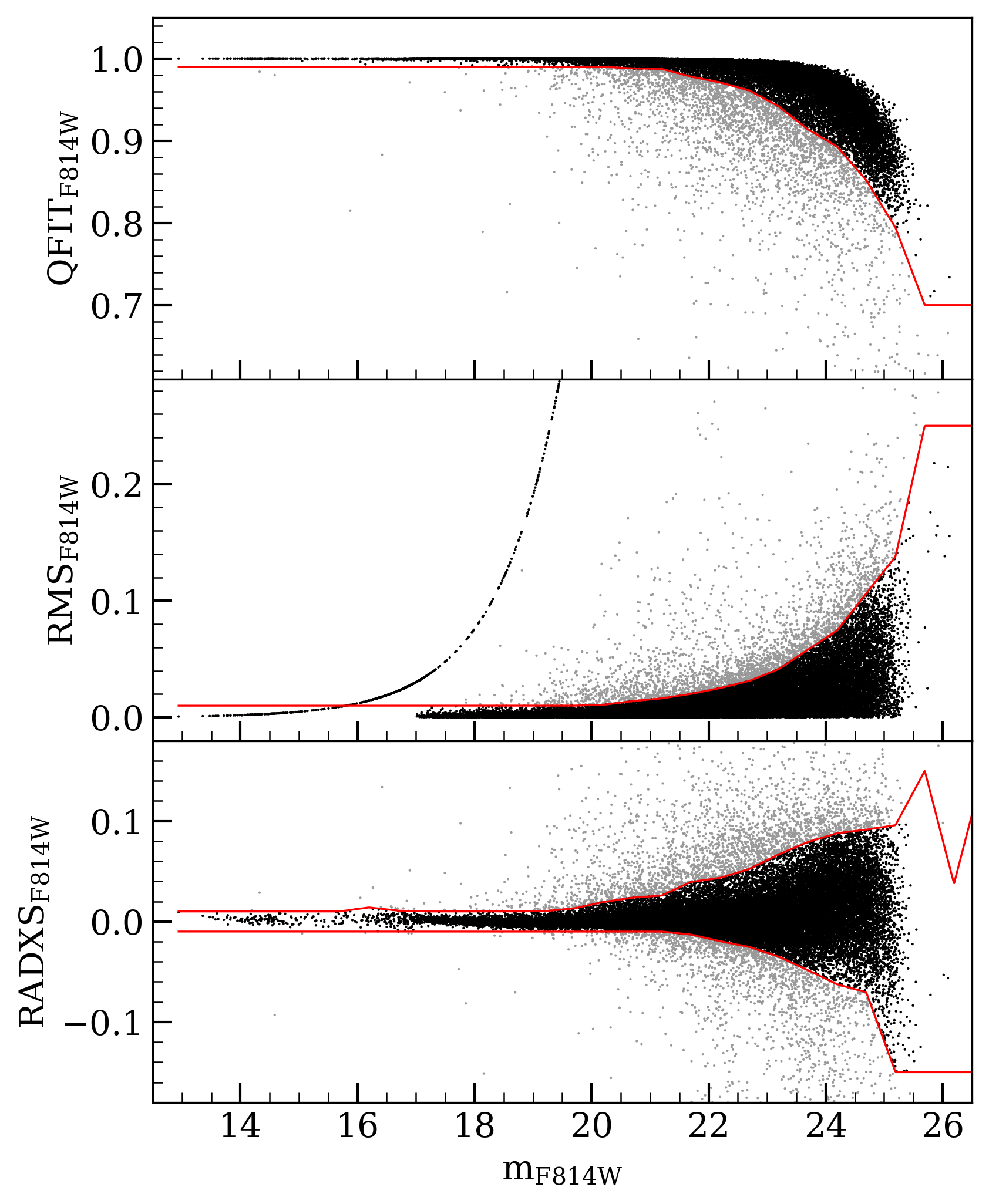} 
\\
\end{tabular}
\caption{Selection of well-measured stars, based on photometric quality parameters, for the cluster NGC~1850. The three sub-plots correspond to observations taken within epoch 3 (2015) in the filters F336W, F438W and F814W (from left to right). Within each sub-plot, the  panels refer to selections based on the \texttt{QFIT} parameter (top panels), the photometric \texttt{RMS} (middle panels) and \texttt{RADXS} parameter (bottom panels). Within each panel the red line follows the magnitude-dependent selection criteria used to separate well-measured stars (black dots) from those with bad quality measurements (grey dots). Stars that are only measured in a single exposure, follow the exponential-like function in the panels showing the \texttt{RMS} distribution. 
\label{fig:phot_quality_sel}
}

\end{figure*}

To correct for these effects, we first defined a sample of well-measured cluster stars as reference objects for the correction. For the selection, we followed the strategy outlined by \citet{Libralato22}. We started by selecting stars for each combination of filter and epoch based on their photometric diagnostics. Fig.~\ref{fig:phot_quality_sel} illustrates the method using the example of NGC~1850 for a set of three different filters. The different panels show the selections based on the \texttt{QFIT} parameter (top panels), the photometric \texttt{RMS} (middle panels) and the \texttt{RADXS} (bottom panels). For each of these three diagnostics, we defined the 95th percentile of their distributions at any magnitude (indicated by the red line in all panels) and selected all objects with values better than this line. In case of the \texttt{QFIT}, better means higher values, whereas for the \texttt{RMS} and \texttt{RADXS} better means smaller (absolute) values. 
For objects that have been measured in only a single exposure, a \texttt{RMS} value obviously cannot be determined. Instead, \texttt{KS2} assigns to their \texttt{RMS} in flux a constant flag, which results into an exponential-like curve when they are transformed into \texttt{RMS} in magnitudes
(see middle panels in Fig.~\ref{fig:phot_quality_sel}). We also flagged these sources as well-measured. Due to uncorrected CTE effects, the distribution of the \texttt{RADXS} parameter as a function of magnitude is skewed towards positive values at fainter magnitudes (see bottom panels in Fig.~\ref{fig:phot_quality_sel}). We therefore opted to determine the 95th percentile curve separately for positive and negative values and selected all sources between those two lines. For each of the three parameters, we additionally set the following conditions: (i) sources with a \texttt{QFIT} better than 0.99 are always kept and objects with a \texttt{QFIT} smaller than 0.7 were always rejected; (ii) sources with a \texttt{RMS} smaller than 0.01 were always kept and objects with an a \texttt{RMS} larger than 0.25 were always discarded; (iii) sources with a |\texttt{RADXS}| smaller than 0.01 were always included while all sources with values larger than 0.15 were always rejected. As additional selection criteria we required the number of good measurements to be at least half the number of the total detections ($\rm N_{\rm g}/\rm N_{\rm f}\ge0.5$) and the fraction of the flux from neighbouring sources within the PSF aperture to be smaller than the flux of the source itself ($o<1$). 
A source qualifies as well-measured if all the above selection criteria are fulfilled in at least two filters (or one filter, in the case the source is only detected in that filter in a given epoch) in at least two epochs. 

Astrometric diagnostics that come from the PM determinations allow us to refine our sample even further to include only stars with reliably-measured PMs. We removed stars for which (i) the reduced $\chi^2$ of the PM fit is larger than 4--5 (depending on the PM quality of a given cluster) in both components, (ii) the fraction of data points actually used for the PM determination is less than 80\%, and (iii) the uncertainty of the PM is greater than the 95th--98th percentile (depending on the cluster) of its distribution at any magnitude.

\begin{figure*}[h]
\centering
\includegraphics[width=2\columnwidth]{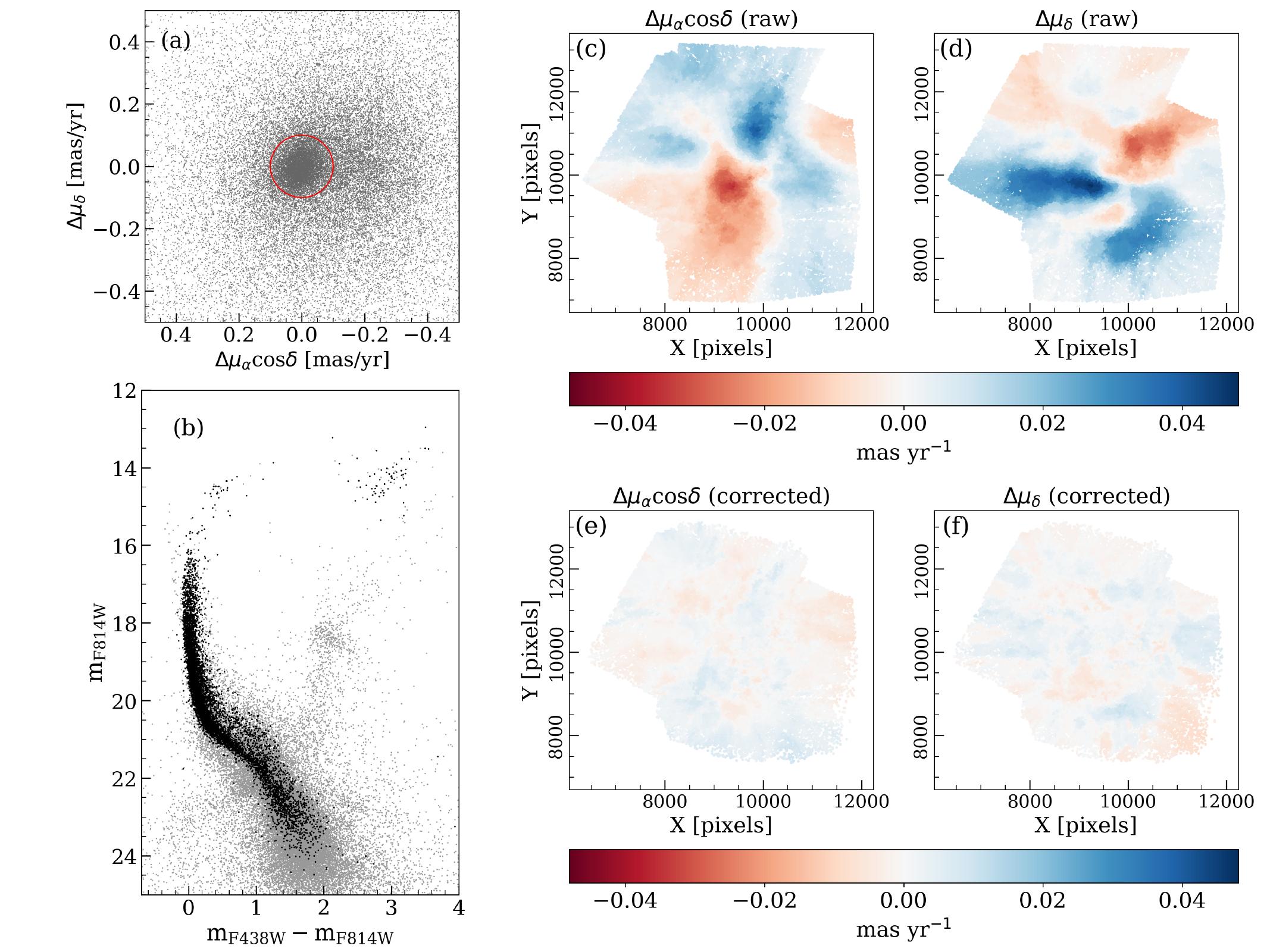}
\caption{Outline of the procedure to correct the PMs for high-frequency systematic effects. Panel (a) shows the VPD of sources towards NGC~1850. Likely cluster members were selected based on their PMs (within the red circle, which has a radius of 0.1~mas\,yr$^{-1}$ in the case of NGC~1850) and their positions in the CMD. Panel (b) shows the $m_{\rm F814W}$ vs $m_{\rm F438W}-m_{\rm F814W}$ CMD of all sources with measured PMs towards NGC~1850 (grey dots) and the final selection of well-measured cluster stars (black dots). Panels (c) and (d) show, separately for each PM component, the maps of the local mean raw PM field (before the high-frequency correction). Each point in the map is colour-coded according to the median PM of its closest 200 well-measured cluster stars. The corresponding maps of the a posteriori corrected PMs are displayed in panels (e) and (f). The points in these maps are coloured according to the same range in colour as in panels (c) and (d).\label{fig:PM_correction}}
\end{figure*}

We finally selected from the sample of well-measured stars those objects that are likely cluster members based on their PMs and their positions in the CMD. In particular, for each cluster we selected a circular region in the VPD that includes the bulk of stars with cluster-like PMs (see panel (a) in Fig.~\ref{fig:PM_correction} for the case of NGC~1850). Then, we defined a region in the CMD that follows the sequence of the cluster stars (see panel (b) in Fig.~\ref{fig:PM_correction} for the $m_{\rm F814W}$ vs $m_{\rm F438W}-m_{\rm F814W}$ CMD of NGC~1850, highlighting selected cluster members). In cases where the cluster region is covered by a number of partially overlapping data sets using different filters (e.g. for NGC~1850), we combined the selections coming from different colour and magnitude combinations.

To correct for the low-frequency systematic effects, we first divided our sample of well-measured cluster stars into different sub-groups, based on the time baseline used for the PM calculation. For each sub-group, we determined the sigma-clipped median of the PM distribution in both directions. By construction, these values should be zero. If any offset is present, we subtracted the determined median PM from all stars in the respective sub-group to shift the PM distribution back to zero.

High-frequency systematic effects can be seen as correlated structures in the maps of the local PMs. Uncorrected CTE and distortion residuals can result in systematic shifts of the median PM, both as a function of magnitude and location on the master frame.
Panels (c) and (d) of Fig.~\ref{fig:PM_correction} show the maps of the locally averaged raw PMs towards NGC~1850 for the $\Delta\mu_{\alpha}\rm cos(\delta)$ and $\Delta\mu_{\delta}$ components, respectively. In the maps, each star is colour-coded by the median PM of the closest 200 well-measured cluster stars. A pattern of PM variations due to the uncorrected high-frequency systematics is clearly evident. To correct for this effect, we followed the methods described in \citet{Bellini14}. Briefly, for each star we selected all well-measured cluster stars within a spatial box of given size (typically 600--750 pixels per side) around the star and within a magnitude interval centred on this star. If there are more reference stars than a pre-defined upper limit (N$_{max}$, typically between 40 and 100), then we selected the closest N$_{max}$ reference stars, both spatially and in magnitude. Then we computed the sigma-clipped median PM (in both directions) of these selected reference stars (excluding the target star itself). By construction, this value should be zero and we corrected the PM of the target star by the determined offset. For the brightest and faintest stars we defined a magnitude threshold for the selection of the reference stars, instead of using a fixed magnitude interval. If the found number of close reference stars for a given star is less than 25, no correction was applied. The exact values of the above mentioned parameters can vary from cluster to cluster. They were empirically determined to give optimal results for the correction, as a compromise between the spatial resolution and the number of reference stars available for the correction of each star. 

The maps of the locally a posteriori corrected PMs are shown in panels (e) and (f) of Fig.~\ref{fig:PM_correction}. The points are coloured according to the same range in colour as in panels (c) and (d). The maps reveal that our corrections were able to significantly mitigate the effects of small-scale systematics. 

We accounted for the uncertainties associated with each step of the correction process in the total error of the a posteriori corrected PMs. 
For both the low-and high-frequency correction, we calculated the standard error of the mean and added these values in quadrature to the uncertainties of the PMs. The a posteriori corrected PMs with the corresponding errors, along with a flag that tells whether a star has been corrected, are reported in the astrometric catalogues of each cluster (see Appendix~\ref{app:catalogues}).

\subsection{Absolute PMs}

Since we used cluster member stars as our reference objects for the transformation of the individual exposures to the master frame, the determined PMs are relative to the bulk motion of the cluster and the PM distribution of cluster stars is centred at zero. To calibrate relative PMs to an absolute scale, traditionally two different methods can be used. The first one involves using background galaxies as reference objects. These galaxies form a non-moving reference frame owing to their large distances and the absolute stellar motions are given by the difference between the relative motions of the stars and the reflex motions of the galaxies. This strategy has been employed to measure the absolute motion of dwarf galaxies \citep[e.g.][]{Sohn13, Sohn17, Sohn20} but also for Galactic open and globular clusters \citep[e.g.][]{Bellini10, Massari13, Libralato18, Libralato18b}. However, in crowded stellar fields, usually only a small number of these background objects can be detected.
Additionally, our PSF-based photometric routine is not optimised for fitting extended sources, such as galaxies, which results in larger positional uncertainties for them. 

\begin{figure}[h]
\centering
\includegraphics[width=\columnwidth]{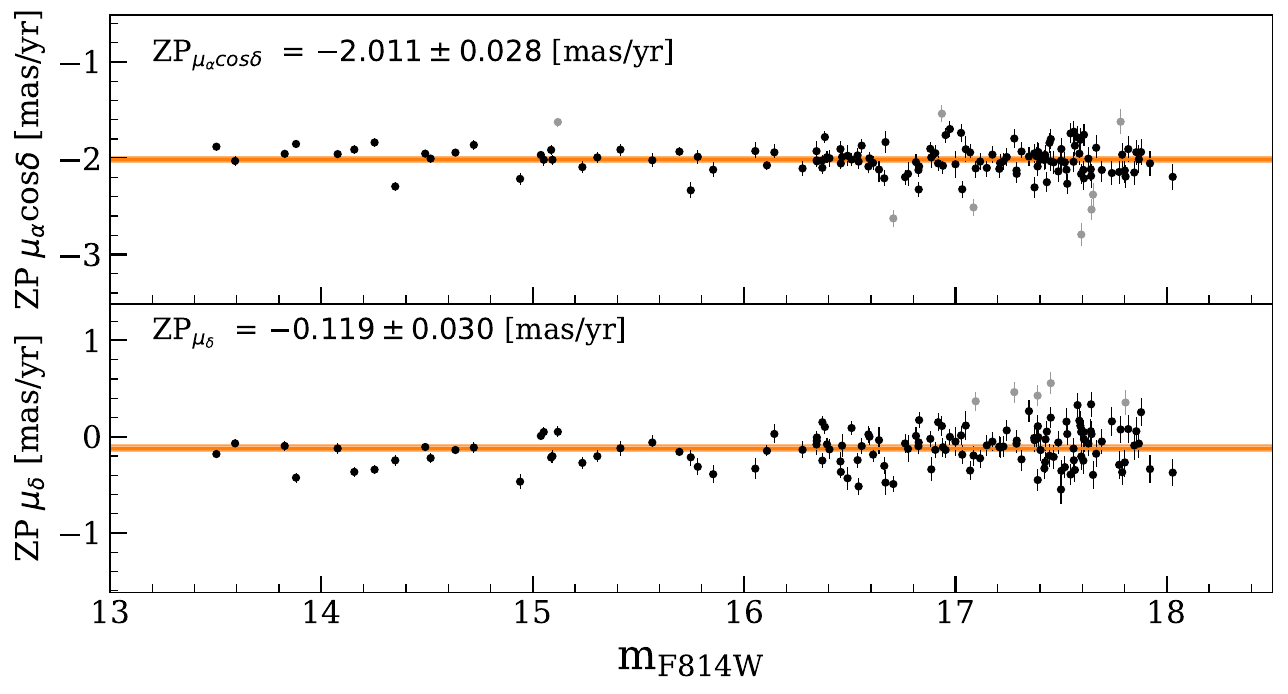}
\caption{Calibration of the relative PMs to an absolute scale using \textit{Gaia} DR3 stars as reference objects. The two panels show the differences between the relative HST PMs and absolute PMs from well-measured \textit{Gaia} stars for NGC~1850. The PM ZP in each direction is determined as the 2.5$\sigma$-clipped median value (orange solid line in each panel). Excluded stars are marked with grey symbols.\label{fig:PM_ZP}}
\end{figure}

Another strategy involves using objects with already known motions to calibrate the PMs to an absolute scale. We are now in the fortunate position that the \textit{Gaia} mission provides us with large numbers of relatively bright objects with measured PMs that can be used for an accurate determination of the PM ZP \citep[see, e.g.][]{Massari21, Libralato22}. Thus, we opted to use \textit{Gaia} DR3 stars for the PM calibration. For each cluster, we started by cross-identifying stars from our high-quality a posteriori corrected sample with the \textit{Gaia} catalogue. In order to include only reference stars that are well-measured in both catalogues, we further applied the following quality selections to the \textit{Gaia} stars: a renormalized unit weight error (ruwe) $<$1.4, an astrometric excess noise $<$ 0.4, a number of bad along-scan observations $<$ 2\% of the total number of along-scan observations and a PM uncertainty $<$0.15~mas\,yr$^{-1}$ in both components. We further excluded stars that are within 15--40~arcsec from the cluster centre (depending on the cluster's size and density) to avoid regions where stellar crowing might be an issue. In the case the above selection criteria result in too few reference stars for a given cluster, we allowed for a more generous cut in the PM uncertainties (up to 0.35~mas\,yr$^{-1}$) in order to have at least 30 stars for the calibration. To determine the ZP of the PMs in both components, we computed the mean difference between the HST and \textit{Gaia} PMs, applying an iterative 2.5$\sigma$-clipping until convergence. 
As an illustration, Fig.~\ref{fig:PM_ZP} shows, using the example of NGC~1850, the difference between the relative HST PMs and the absolute \textit{Gaia} PMs for the selected sample of reference stars. The orange solid line in both panels corresponds to the adopted ZP for $\mu_{\alpha}\rm cos\delta$ and $\mu_{\delta}$, respectively.
The uncertainty of the calibration is composed of two terms that we summed in quadrature to get the total ZP uncertainty: the first is the statistical error, given by the error of the mean of the distribution of the HST--\textit{Gaia} PM differences. The second term refers to the systematic uncertainty of the \textit{Gaia} DR3 PMs of 0.026~mas\,yr$^{-1}$ \citep[see][]{Vasiliev21, Libralato22}. 
We remind the reader here that the distribution of the relative PMs of cluster stars is centred at zero. Thus, by construction, the absolute motion of a given cluster directly corresponds to the negative value of the ZPs in RA and Dec directions.

\begin{figure*}[h]
\centering
\includegraphics[width=1.7\columnwidth]{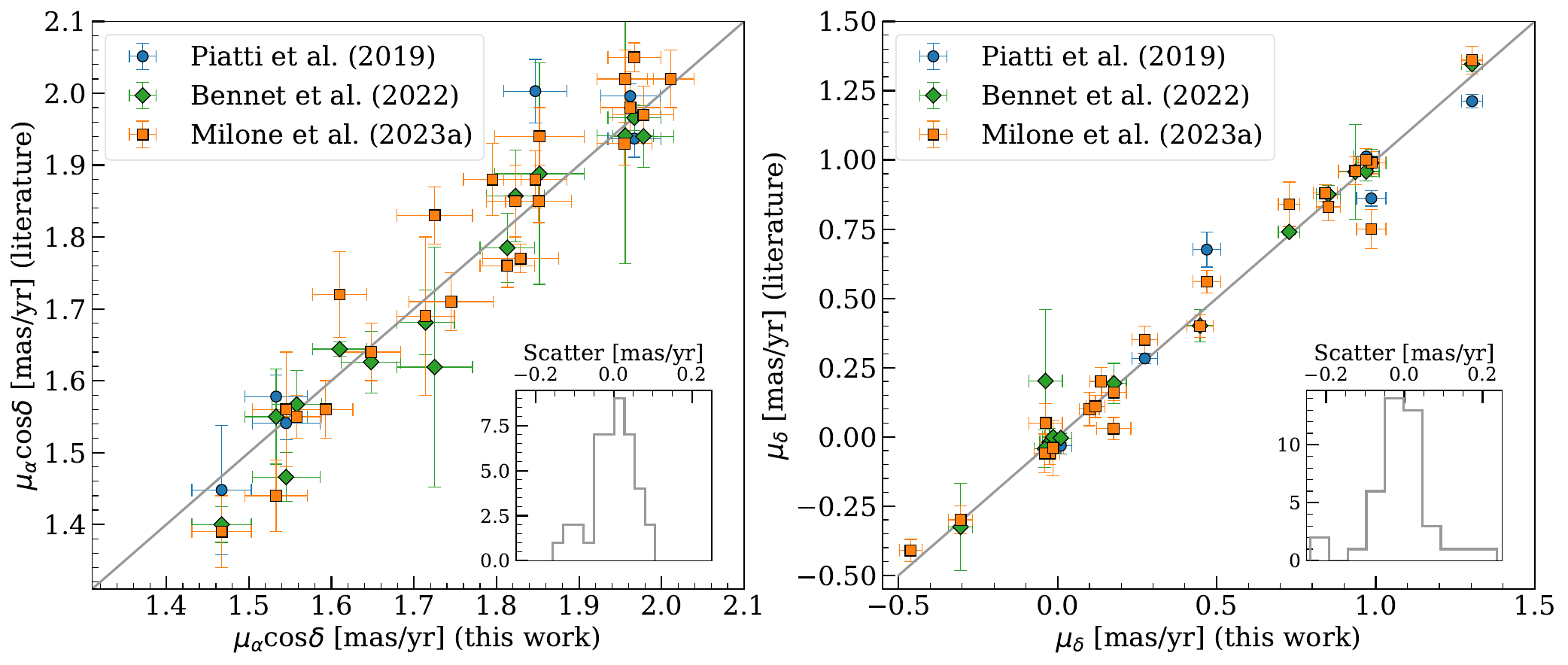}
\caption{Comparison between the HST-based absolute motions of the clusters presented in this work and measurements from the literature. The grey lines follow a one-to-one relation and are not a fit to the data. The insets within the two panels show the distribution of the points perpendicular to the grey line, illustrating the differences in the measurements. }\label{fig:PM_compar}
\end{figure*}

The ZPs for all clusters, along with the corresponding uncertainties, are reported in Table~\ref{tab:PM_ZP}, and also written in the headers of the astrometric catalogues of each cluster (see Appendix~\ref{app:catalogues}). In Fig.~\ref{fig:PM_compar} we also present a comparison between the absolute motions of the clusters derived in this work and literature measurements from \cite{Piatti19}, \citet{Bennet22} and \citet{Milone23a}. The shown points closely follow a one-to-one relation, indicating a good agreement. To estimate the typical difference between our results and the literature measurements, we calculated for each cluster the distance perpendicular to the one-to-one relation. The distribution of the distances is shown in the insets within the two panels of Fig.~\ref{fig:PM_compar}. We found a standard deviation of 0.06~mas\,yr$^{-1}$ for the distribution along the $\mu_{\alpha}\rm cos\delta$ direction and a standard deviation of 0.08~mas\,yr$^{-1}$ for the distribution along the $\mu_{\delta}$ direction. 
We further examined whether there are any systematic trends between the literature values and our measurements. To do so, we fitted a least-squares straight line to the points shown in Fig.~\ref{fig:PM_compar}. We did this separately for each of the three literature studies. We found no significant deviation from a one-to-one relation. There might only be a slight trend when comparing our results with the measurements from \citet{Milone23a} along the $\mu_{\alpha}\rm cos\delta$ direction, in the sense that the values as determined by \citet{Milone23a} tend to be larger than the ones from this study at the upper end of the distribution. However, this is only significant at the 1$\sigma$ level.

\subsection{Absolute astrometry\label{subsec:astrometry}}

For each cluster, we used the cross-match between our high-quality PM catalogues and the $Gaia$ DR3 catalogue to register the X and Y master-frame coordinates to an absolute astrometric frame. We employed the \texttt{astropy} module \texttt{fit\_wcs\_from\_points} to find the best-fitting WCS solution that we used to project the X and Y coordinates of all stars to RA and Dec. The celestial positions are given in the ICRS at the epoch 2016.0, the reference epoch of \textit{Gaia} DR3.


\section{Demonstration: PMs towards NGC~1850}\label{sec:NGC1850_demo}

\begin{figure*}[h]
\centering
\includegraphics[width=2\columnwidth]{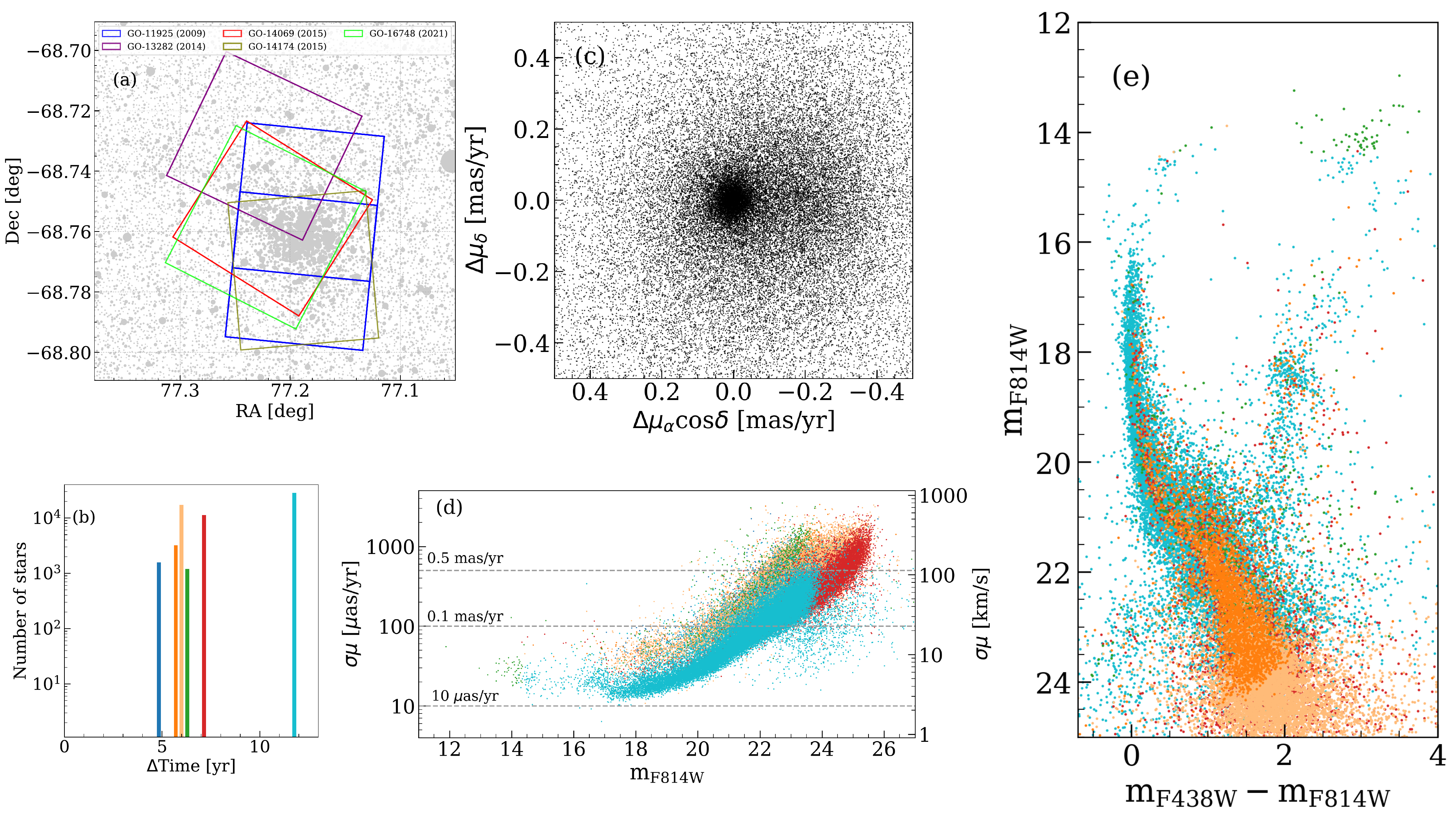}
\caption{Outline of the measured PMs towards NGC~1850. (a) Sketch of HST footprints from different epochs used for the calculation of the PMs. (b) Histogram of the time baselines available for the PM calculations of each source. The range in time baselines is mainly due to the varying overlap of the observations from different epochs. Sources shown in panels (d) and (e) are colour-coded according to the time baselines shown here. (c) VPD of a posteriori corrected, relative PMs. Member stars of NGC 1850 are clustered around (0.0, 0.0). (d) Corrected 1D PM errors as a function of the $m_{\rm F814W}$ magnitude. Bright well-measured stars reach a precision of $\sim$15~$\mu$as\,yr$^{-1}$ (corresponding to $\sim$3.3~km\,s$^{-1}$ assuming a distance of 47.4~kpc). (e) $m_{\rm F814W}$ vs $m_{\rm F438W}-m_{\rm F814W}$ CMD of all sources with measured PMs.   \label{fig:ngc1850_PM}}
\end{figure*}

\begin{figure*}
\begin{tabular}{cc}
\includegraphics[width=\columnwidth]{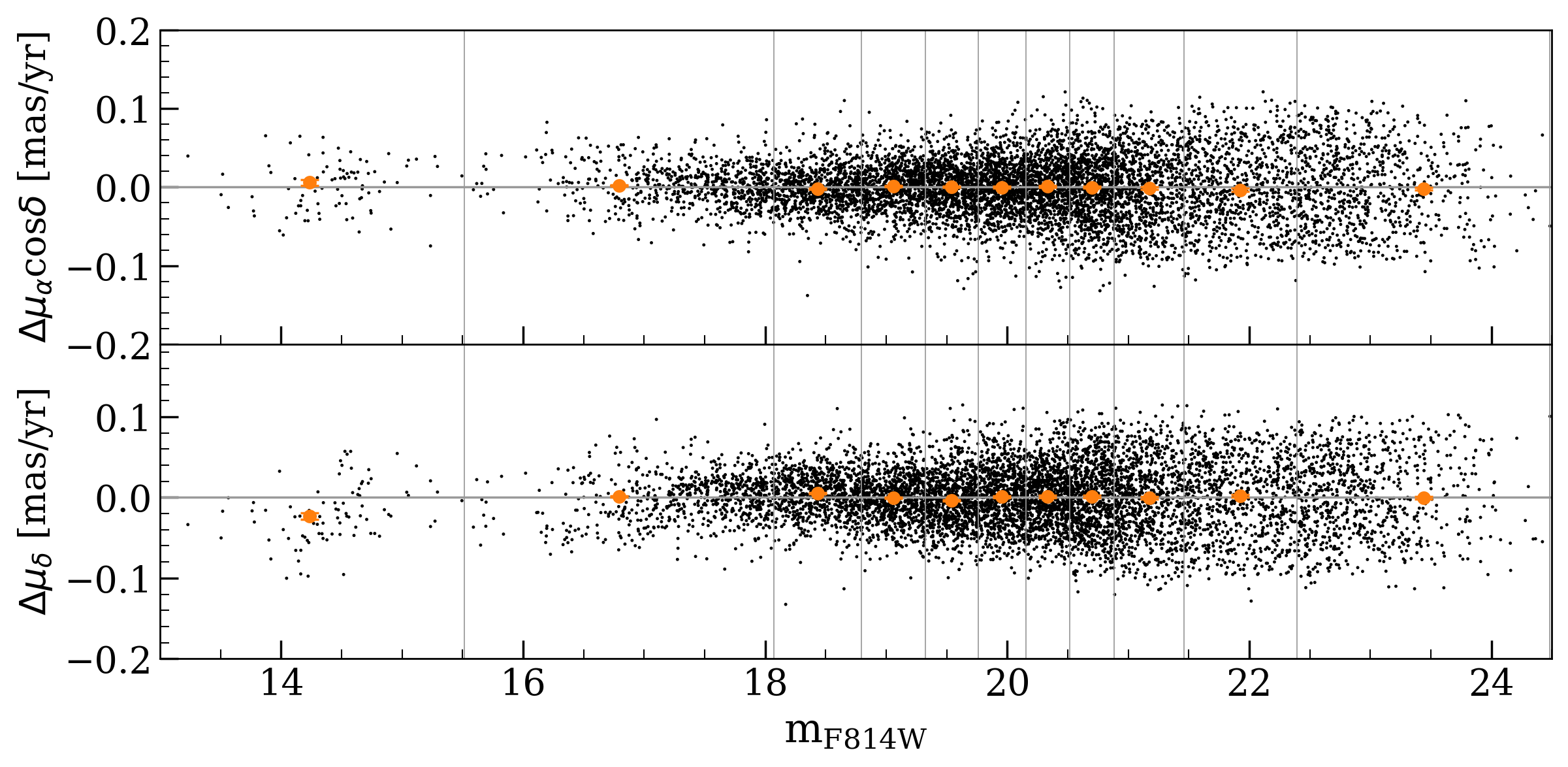} &
\includegraphics[width=\columnwidth]{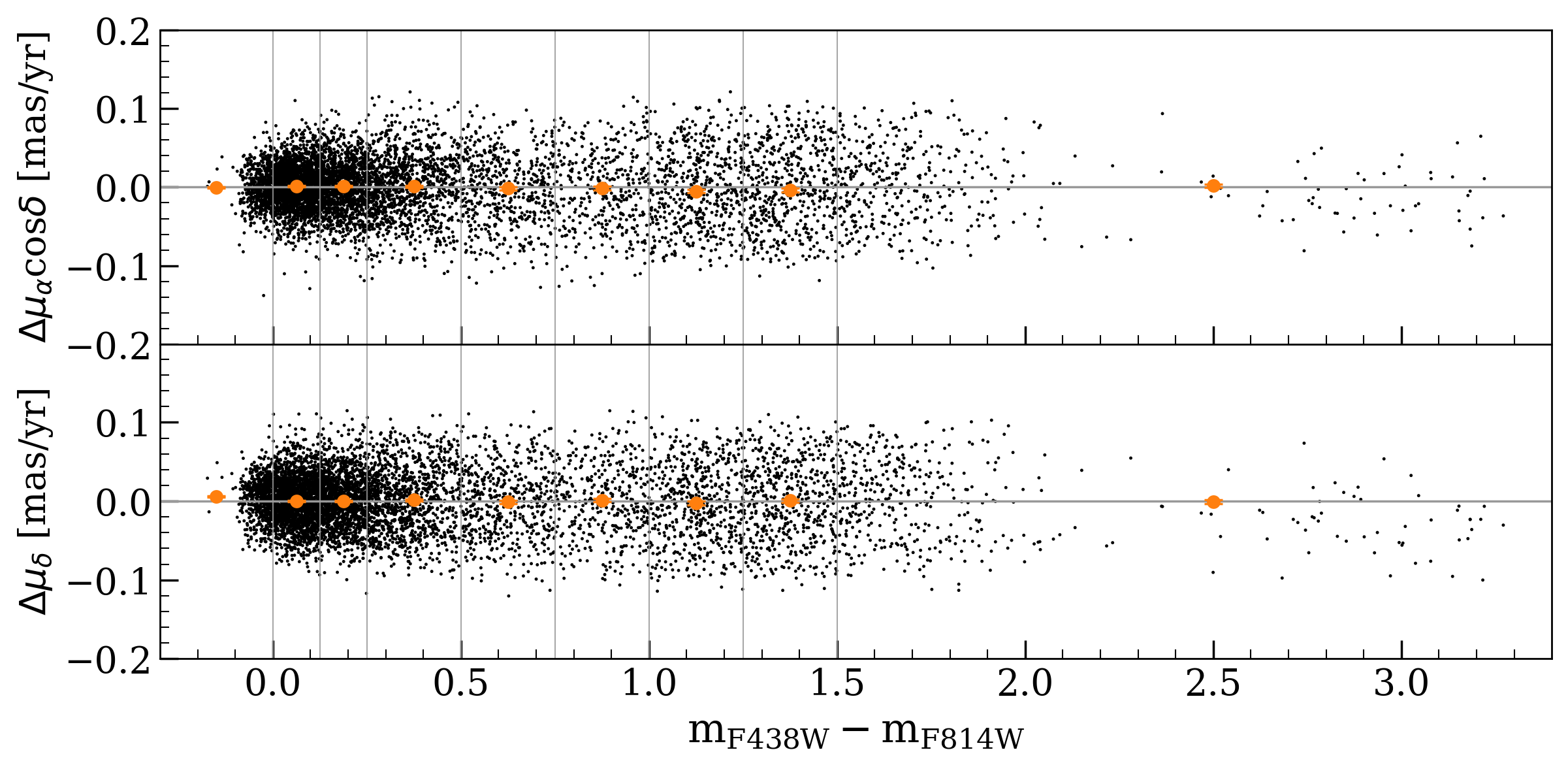} \\
\end{tabular}
\caption{Relative a posteriori corrected PMs, $\Delta\mu_{\alpha}\rm cos\delta$ and $\Delta\mu_{\delta}$, of well-measured NGC~1850 member stars as a function of the $m_{\rm F814W}$ magnitude (left panels) and the $m_{\rm F438W}-m_{\rm F814W}$ colour (right panels). The orange points are median values within bins containing approximately equal numbers of objects (edges of the bins are indicated as grey vertical lines). Error bars, calculated as the standard error of the mean, are also shown but are generally smaller than the shown points. The horizontal grey line in each panel is set at $\Delta\mu=0.0$~mas\,yr$^{-1}$. 
\label{fig:ngc1850_pm_vs_mag}
}

\end{figure*}

To showcase some applications of our PM data set, we present here the analysis of the PMs towards the cluster NGC~1850. This is to demonstrate the range of possibilities and limitations of the current PM data set. 
NGC~1850 is a young ($\sim$70--100~Myr) cluster and is located at the western edge of the LMC bar. Its CMD shows an extended main sequence turn-off \citep{Bastian16} and split upper main sequence \citep{Correnti17,Milone18}. Both phenomena are common features in young massive star clusters in the Magellanic Clouds and are most likely caused by stars that rotate at different rates \citep[see, e.g.][]{Niederhofer15, Kamann23}. Fig.~\ref{fig:ngc1850_PM} shows an overview of the PM catalogue of NGC~1850. For this cluster, we used WFC3/UVIS data sets from five different HST programmes. These observations have been taken between 2009 and 2021. The field of view covered by the different programmes is shown in panel (a). The different overlaps of the individual observations result in time baselines available for the PM determination between $\sim$4.7 and $\sim$11.9~yr. Panel (b) shows the distribution of the time baselines. The VPD of the relative, a posteriori corrected PMs is displayed in panel (c). 
The corresponding 1D uncertainties of the corrected PMs as a function of the F814W magnitude are shown in panel (d). The points are coloured according to the histogram of time baselines from panel (b). Bright, well-measured sources reach a precision of $\sim$15~$\mu$as\,yr$^{-1}$ (including the uncertainties from the a posteriori correction), corresponding to $\sim$3.3~km\,s$^{-1}$, assuming a distance to the cluster of 47.4~kpc (see Table~\ref{tab:cluster_param}). A CMD of stars with measured PMs is presented in panel (e). Also here, the sources are coloured according to the histogram shown in panel (b). The final catalogue for NGC~1850 contains 62\,984 sources with measured PMs. We checked the final corrected PMs for any residual magnitude- or colour-dependent systematic trends. Fig.~\ref{fig:ngc1850_pm_vs_mag} shows the relative, a posteriori corrected PMs, $\Delta\mu_{\alpha}\rm cos\delta$ and $\Delta\mu_{\delta}$, of well-measured cluster stars (see Section~\ref{subsec:PM_corr}) as a function of the $m_{\rm F814W}$ magnitude and the $m_{\rm F438W}-m_{\rm F814W}$ colour, respectively. In all panels, the orange points denote the median PMs, calculated in bins containing approximately equal numbers of sources. From the figure, no clear systematic trend of the corrected PMs is evident, neither as a function of magnitude nor as a function of colour.

For the following demonstration cases, we applied several quality criteria to the PM catalogue of NGC~1850 in order to keep only sources with the best measurements. We refer to this sample as the ``high-quality'' sample. We started by applying the photometric and astrometric quality cuts as described in Section~\ref{subsec:PM_corr}. Then, we used additional constraints on the PM uncertainties. In particular, we retained only sources for which (i) the error on the PM is smaller than the 90th percentile at any magnitude and (ii) the error on the PMs in both components is smaller than 50~$\mu$as\,yr$^{-1}$. The so-selected, final list contains a total of 8\,250 high-quality sources. 

\subsection{Dynamics of different stellar populations towards NGC~1850\label{subsec:stellar_pop_dyn}}

\begin{figure}[h]
\centering
\includegraphics[width=\columnwidth]{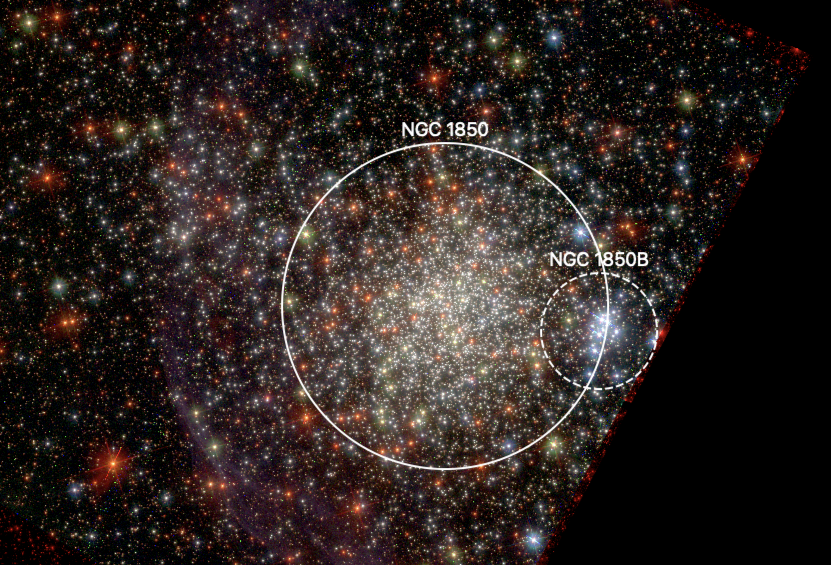}
\caption{Three-colour image of NGC~1850, created from HST observations in the F336W (blue), F438W (green) and F814W (red) filters. Member stars of
NGC~1850 are selected based on their motion and distance
from the cluster centre (white solid circle). Stars belonging to
the young group called NGC~1850B are selected as the young
sequence in the CMD that are located within the white dashed
circle and show consistent motions. \label{fig:ngc1850b}}
\end{figure}

\begin{figure*}
\begin{tabular}{cc}
\includegraphics[width=0.95\columnwidth]{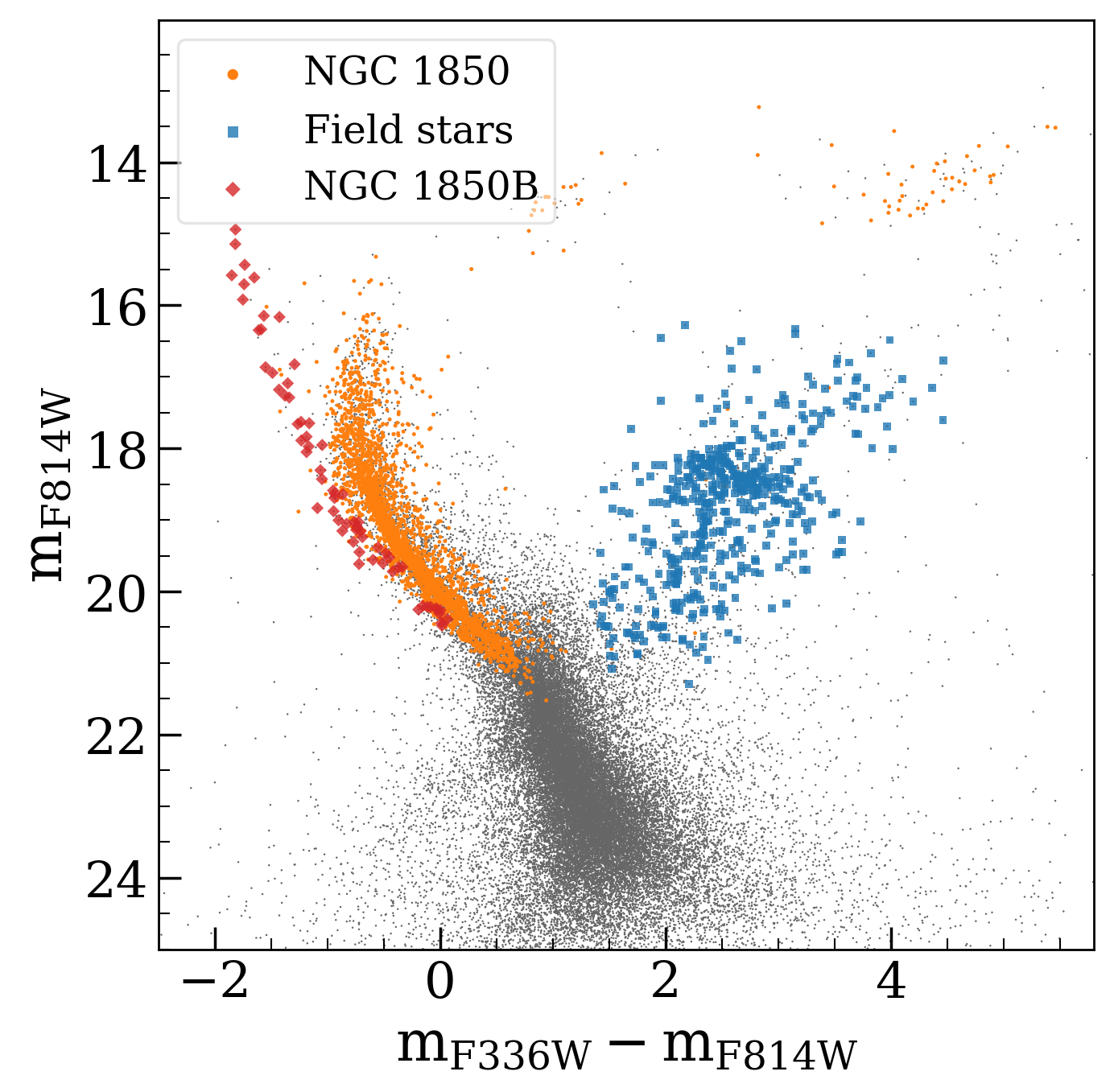} &
\includegraphics[width=\columnwidth]{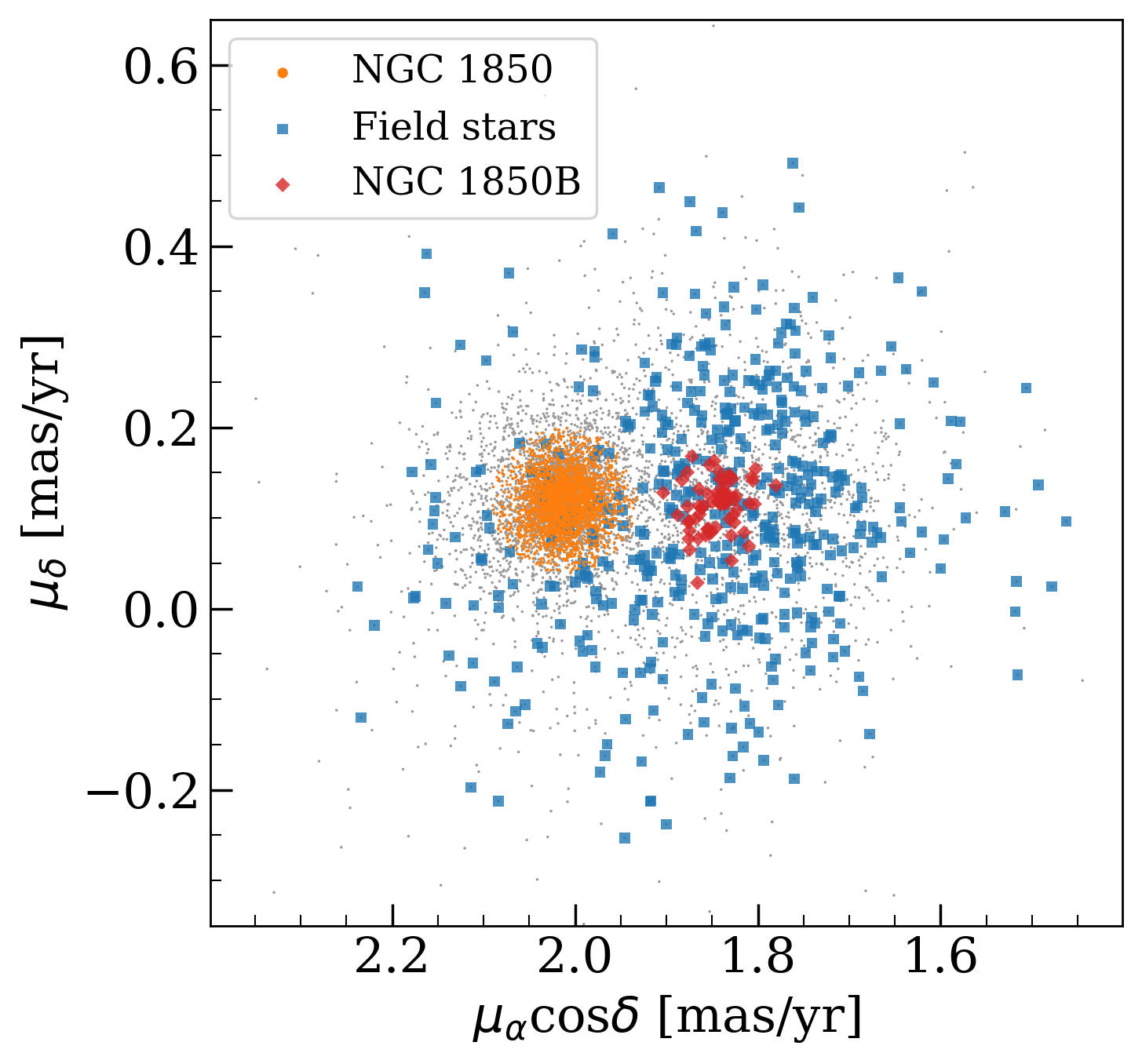} \\
\end{tabular}
\caption{The different populations in the field of NGC~1850. \textit{Left}: $m_{\rm F814W}$ vs $m_{\rm F336W}-m_{\rm F814W}$ CMD highlighting the selected members of NGC~1850 (orange dots), NGC~1850B (red diamonds) and old field stars (blue squares). \textit{Right}: VPD of the same populations. 
\label{fig:ngc1850_populations}
}

\end{figure*}

Here, we analyse the kinematics of the various stellar populations that are present within our field of view containing NGC~1850. As stars belonging to the cluster, we selected all stars that are within 35~arcsec ($\sim$1.5$\times$ its scale radius, see Table~\ref{tab:cluster_param} and Fig.~\ref{fig:ngc1850b}) from the cluster centre, and have relative PMs smaller than 80~$\mu$as\,yr$^{-1}$, corresponding to $\sim$3 times its standard deviation.
The left panel of Fig.~\ref{fig:ngc1850_populations} shows the $m_{\rm F814W}$ vs $m_{\rm F336W}-m_{\rm F814W}$ CMD of all stars with measured PMs towards NGC~1850. The PM-cleaned sample of cluster stars is highlighted as orange points in the diagram. Our quality and PM selections trace the features of the cluster, including the evolved stars, the extended main sequence turn-off and the split main sequence, down to $m_{\rm F814W}\sim21$~mag. The VPD of the cluster members is shown in the right panel of Fig.~\ref{fig:ngc1850_populations}. Note that we show here the absolute motions. Since the relative PM of the cluster is zero, the absolute bulk motion of NGC~1850 is simply given by the negative values of the corresponding PM ZPs in both components. We found:
\begin{equation*}
    \begin{aligned}
    & \mu_{\alpha}\rm cos\delta = 2.011\pm0.028~\mathrm{mas}\,\mathrm{yr}^{-1}, & \mu_{\delta} = 0.119\pm0.030~\mathrm{mas}\,\mathrm{yr}^{-1},
    \end{aligned}
\end{equation*}
in good agreement with recent measurements from \citet{Milone23a, Milone23b}.

Close to NGC~1850, there is a compact group of very young stars, dubbed NGC~1850B. This small cluster, which has an age of $\sim$5--15~Myr \citep{Gilmozzi94, Sollima22} is located about 30~arcsec (which corresponds to a projected physical separation of $\sim$7~pc at the distance of the LMC) west of NGC~1850 (see Fig.~\ref{fig:ngc1850b}) and presumably lies at a similar distance as its larger neighbour \citep[e.g.][]{Milone23a}. Therefore, NGC~1850 is sometimes considered to be a double cluster in the literature \citep[see, e.g.][and references therein]{Sollima22}. 
In particular, this means that the clusters are physically associated and might merge in the future. While binary clusters are quite common in the LMC \citep[see, e.g.][and references therein]{Dieball02, Mucciarelli12} they normally have very similar ages and metallicities. In this respect, if NGC~1850 was a double cluster, this would depict a very special case, given the large difference in age.

We used our PMs to disentangle the motions of these two objects. To define stars that belong to NGC~1850B, we first selected all stars that are within a circle of 16~arcsec centred at the young cluster (indicated by the dashed white circle in Fig.~\ref{fig:ngc1850b}). We further considered only stars that follow the young sequence in the $m_{\rm F814W}$ vs $m_{\rm F336W}-m_{\rm F814W}$ CMD and show consistent PMs. Our final sample of stars belonging to NGC~1850B is illustrated as red diamonds in both panels of Fig.~\ref{fig:ngc1850_populations}. From the VPD it is immediately clear that NGC~1850B has a PM that is distinctive from that of NGC~1850 \citep[see also appendix in the study from][]{Milone23b}. We determined the bulk motion of the young group as the sigma-clipped median value of the selected member stars. This yields:
\begin{equation*}
    \begin{aligned}
    & \mu_{\alpha}\rm cos\delta = 1.843\pm0.028\pm0.003~\mathrm{mas}\,\mathrm{yr}^{-1}\\
    & \mu_{\delta} = 0.116\pm0.030\pm0.003~\mathrm{mas}\,\mathrm{yr}^{-1}.
    \end{aligned}
\end{equation*}
Here, the two uncertainties quoted for each value correspond to the error in the PM ZP and the statistic uncertainty from the determination of the median.
To determine the full motions of the two clusters in three dimensions, we supplemented our PM measurements with the MUSE spectroscopic catalogue of NGC~1850, presented by \citet{Kamann23}. This catalogue contains line-of-sight velocity measurements of 4\,200 stars within two MUSE pointings. We cross-identified stars in common with our high-quality PM sample and found a value of $v_{\rm NGC1850B}=252.4 \pm 1.4$~km\,s$^{-1}$ as the line-of-sight velocity of NGC~1850B, resulting from the sigma-clipped median of our PM selected member stars. For the main cluster, we adopted a value of $v_{\rm NGC1850}=247.1 \pm 0.2$~km\,s$^{-1}$, as determined by \citet{Kamann23}. These values demonstrate that the difference in motion between the two clusters is mainly along the RA direction, whereas the velocities along the line-of-sight and in Dec direction are similar. 

To evaluate the velocities in a more descriptive way, we transformed the PMs and line-of-sight velocities into the reference frame of the LMC. This frame is centred on the dynamical centre of the galaxy and is orientated such that the x-y plane is aligned with the disc of the LMC and the z-axis is perpendicular to that plane. We projected the observed motions into the frame of the LMC using the formalism outlined in detail in \citet{vanderMarel01b} and \citet{vanderMarel02}. For the transformation, a knowledge of the centre position, distance and orientation of the LMC disc plane is needed, as well as the centre-of-mass motion of the galaxy in the plane of the sky and in line-of-sight direction. For the centre position, viewing angles and PMs, we assumed the values as determined by \citet{Niederhofer22}. We adopted a distance to the LMC of 49.9~kpc \citep{deGrijs14} and a line-of-sight velocity of 262.2~km\,s$^{-1}$ \citep{vanderMarel02}. For the positions of the clusters within the LMC, we assumed the distance and coordinates of NGC~1850 as given in Table~\ref{tab:cluster_param} and shifted the position by 30~arcsec towards the west, to obtain the coordinates of NGC~1850B.

\begin{table}\small
\centering
\caption{Velocity components of NGC~1850 and NGC~1850B projected into the reference frame of the LMC\label{tab:velocities_ngc1850ab}}
\begin{tabular}{lccc}
\hline\hline
\noalign{\smallskip}
Cluster & $v_r$ & $v_{\phi}$ & $v_z$\\
& [km\,s$^{-1}$] & [km\,s$^{-1}$] & [km\,s$^{-1}$] \\
\noalign{\smallskip}
\hline
\noalign{\smallskip}
NGC 1850 & $-$19.1$\pm$6.1 & $-$9.9$\pm$6.0 & 15.7$\pm$3.7 \\
NGC 1850B & $-$22.1$\pm$6.1 & 26.5$\pm$6.0 & 26.3$\pm$3.8 \\
\noalign{\smallskip}
\hline

\end{tabular}

\end{table}

The resulting velocity components in cylindrical coordinates ($v_r$, $v_\phi$ and $v_z$) for the two clusters are reported in Table~\ref{tab:velocities_ngc1850ab}. 
We note here that the exact values of these quantities depend on the adopted parameters of the LMC, for which a variety of determinations exist in the literature. Their specific choice, however, has no effect on our final conclusions.
The velocities of both objects show a similar negative radial component, meaning that the motions deviate from circular orbits and are orientated more towards the centre of the LMC. Additionally, both clusters move perpendicular to the disc of the galaxy, in the positive z-direction, with NGC~1850B having a higher velocity. The largest discrepancy in the kinematics of the two clusters is the tangential velocity $v_{\phi}$. The young group NGC~1850B has a rotational velocity about the centre of the LMC that is in good agreement with the rotation curve of the field stellar population of the galaxy \citep[see, e.g][]{vanderMarel14, Gaia21, Niederhofer22}. The main cluster NGC~1850, however, seems to have a slight negative tangential velocity. This disagreement in $v_{\phi}$ suggests that both clusters are dynamically not related, as also suggested by \citet{Milone23b}, and their proximity might largely be a projection effect.


\subsection{Velocity dispersion}

In this section, we use the measured PMs to estimate the velocity dispersion of the old field star population and of NGC~1850. We started by selecting a clean sample of LMC field stars from our high-quality catalogue. We identified field stars along the red giant branch (RGB) and red clump in the $m_{\rm F814W}$ vs $m_{\rm F438W}-m_{\rm F814W}$ CMD and additionally constrained this selection to stars with distances larger than 40~arcsec from the centre of NGC~1850. We decided for this option over selecting LMC stars based on their PMs, since their distribution in the VPD significantly overlaps with the clump of cluster stars. Our final sample of LMC stars is shown as blue square symbols in both panels of Fig.~\ref{fig:ngc1850_populations}. 

To determine the velocity dispersion in the plane of the sky, we employed a maximum-likelihood approach introduced by \citet{vanderMarel10}, which takes into account the uncertainties of the PM measurements of the individual sources \citep[see also][]{Raso20, Libralato22}. For this, we constructed the following log-likelihood function:
\begin{equation}
\begin{aligned}
\mathrm{ln} \mathcal{L} = -0.5  \sum\limits_{i=1}^n & \Bigg(\mathrm{ln}\big( \sigma_{\mu}^2 + \epsilon_{\mu_{\alpha}\rm cos\delta,i}^2\big) +\frac{\big(\mu_{\alpha}\rm cos\delta_{,i} - \mu_{\alpha}\rm cos\delta_{,0}\big)^2}{\sigma_{\mu}^2 + \epsilon_{\mu_{\alpha}\rm cos\delta,i}^2} \\
+ & \mathrm{ln}\big(\sigma_{\mu}^2 + \epsilon_{\mu_{\delta,i}}^2\big) +\frac{\big(\mu_{\delta,i} - \mu_{\delta, 0}\big)^2}{\sigma_{\mu}^2 + \epsilon_{\mu_{\delta,i}}^2} \Bigg).
\end{aligned}
\end{equation}
In this equation, $\mu_{\alpha}\rm cos\delta_{,i}$ and $\mu_{\delta,i}$ correspond to the measured PMs of the individual sources in RA and Dec direction, whereas $\epsilon_{\mu_{\alpha}\rm cos\delta,i}$ and $\epsilon_{\mu_{\delta,i}}$ are the corresponding uncertainties. Further, $\mu_{\alpha}\rm cos\delta_{,0}$ and $\mu_{\delta, 0}$ denote the RA and Dec components of the bulk motion of the stellar population, and $\sigma_{\mu}$ is the velocity dispersion. We explored the posterior probability distribution using again the MCMC sampler \texttt{emcee}. We set up an ensemble of 50 walkers and ran the MCMC for 5000 steps, considering only the last 25 per cent of the chain. The final values and the corresponding uncertainties result from the 50th, 16th, and 84th percentiles of the final marginalized distributions. 

For the bulk motion of the RGB field stars at the location of NGC~1850 this yields:
\begin{equation*}
    \begin{aligned}
    & \mu_{\alpha}\rm cos\delta = 1.852\pm0.028\pm0.006~\mathrm{mas}\,\mathrm{yr}^{-1} \\
    & \mu_{\delta} = 0.107\pm0.030\pm0.006~\mathrm{mas}\,\mathrm{yr}^{-1},
    \end{aligned}
\end{equation*}
where the second contribution to the associated uncertainties comes from the posterior probability distribution. Assuming the same parameters of the LMC's orientation, distance, line-of-sight velocity and position of the dynamical centre as in Section~\ref{subsec:stellar_pop_dyn}, this results in a centre-of-mass motion of the LMC of: 
\begin{equation*}
    \begin{aligned}
    & \mu_{\alpha}\rm cos\delta = 1.871\pm0.028\pm0.006~\mathrm{mas}\,\mathrm{yr}^{-1} \\
    & \mu_{\delta} = 0.291\pm0.030\pm0.006~\mathrm{mas}\,\mathrm{yr}^{-1}.
    \end{aligned}
\end{equation*}
This inferred value is in good agreement with the range of different measurements from the recent literature \citep[e.g.][]{vanderMarel14, Wan20, Gaia18, Gaia21, Niederhofer22}.

For the velocity dispersion of RGB field stars that are located at the western edge of the LMC bar, we found a value of $\sigma_{\mu} = 0.128 \pm 0.003$~mas\,yr$^{-1}$. Assuming a distance of 49.9~kpc, this translates to $30.3 \pm 0.7$~km\,s$^{-1}$. To put our derived value of the velocity dispersion of RGB stars into context, we compare them to results from the literature. Our PM-based value is larger than what was inferred from line-of-sight velocities. Recently, \cite{Kamann23} and \cite{Song21} used spectroscopic data to study NGC~1850. For the velocity dispersion of the surrounding field star population, they found values of $20.2 \pm 0.8$~km\,s$^{-1}$ and 23.6$^{+1.7}_{-2.6}$~km\,s$^{-1}$, respectively. \cite{vanderMarel14} combined HST PMs with line-of-sight velocities of different stellar populations to study the velocity field of the LMC. For their sample of old stars, they found a line-of-sight velocity dispersion of 22.8~km\,s$^{-1}$. In contrast, our result is in good agreement with what was previously found based on PM measurements. \citet{Choi21} used a compilation of PM data from \textit{Gaia} eDR3 \citep{Gaia21b} together with line-of-sight velocity measurements of RGB, asymptotic giant branch and red supergiant stars to model the three-dimensional kinematics of the LMC stars. They inferred a PM dispersion in RA direction of $0.124 \pm 0.002$~mas\,yr$^{-1}$ ($29.4 \pm 0.5$~km\,s$^{-1}$) and in Dec direction of $0.138 \pm 0.002$~mas\,yr$^{-1}$ ($32.7 \pm 0.5$~km\,s$^{-1}$). Recently, \citet{Libralato23} derived stellar PMs within the central region of the LMC, combining data of the calibration field of the \textit{James Webb} Space Telescope (JWST) Near Infrared Imager and Slitless Spectrograph (NIRISS) with archival HST observations. They reported a velocity dispersion of RGB stars of $0.144 \pm 0.003$~mas\,yr$^{-1}$ ($33.8 \pm 0.6$~km\,s$^{-1}$). Taken together, all these measurements hint towards an anisotropic velocity field, in the sense that the velocity dispersion of the stars is larger within the plane of the LMC disc (which is measured by the PMs, since we see the LMC almost face-on) than perpendicular to it (which, in turn, is traced by the line-of-sight velocities).

\begin{figure}[h]
\centering
\includegraphics[width=\columnwidth]{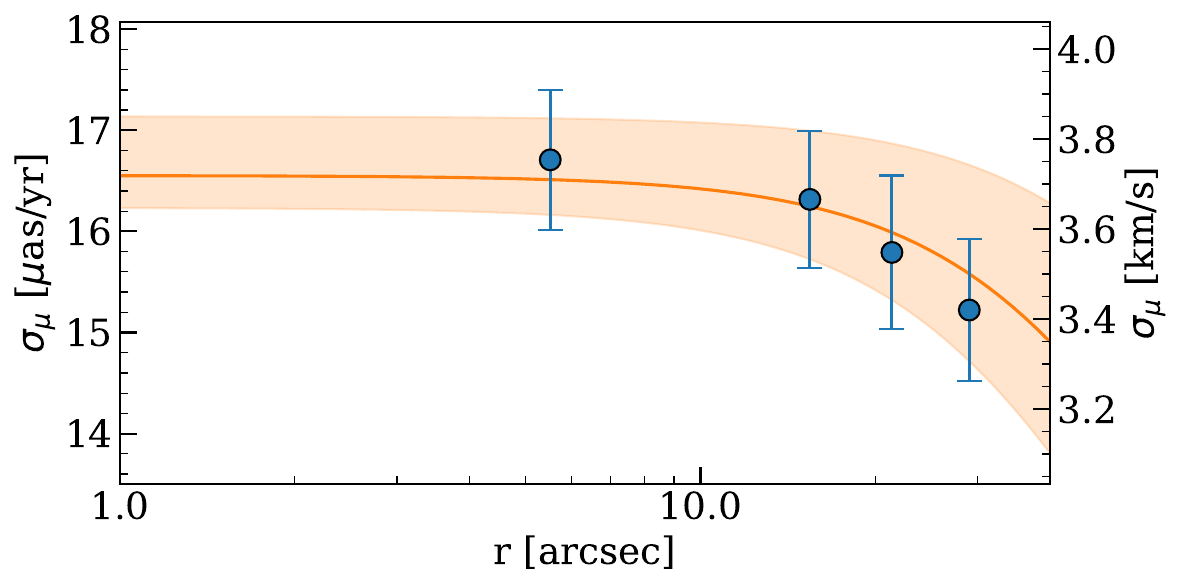}
\caption{Velocity dispersion profile of NGC~1850 in the plane of the sky, obtained from our PM measurements (blue dots with errorbars). The best-fit Plummer profile to the discrete stellar PMs is shown as an orange solid line. The shaded region corresponds to the uncertainties in the parameters of the fit.  \label{fig:ngc1850_vel_disp_pm}}
\end{figure}

\begin{figure}[h]
\centering
\includegraphics[width=\columnwidth]{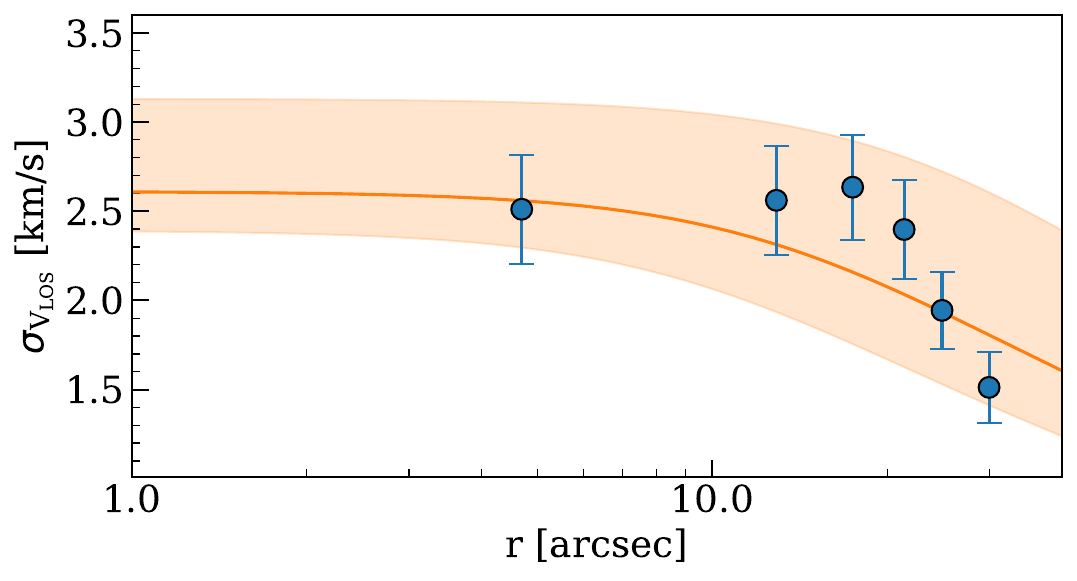}
\caption{Similar to Fig.~\ref{fig:ngc1850_vel_disp_pm}, but now for the line-of-sight velocity dispersion obtained from MUSE data. \label{fig:ngc1850_vel_disp_rv}}
\end{figure}

We used our PM catalogue to provide a first estimate of the intrinsic global velocity dispersion in the plane of the sky of NGC~1850, as well as its radial profile. For the determination of the velocity dispersion, we employed again the maximum-likelihood method described above. We further refined our high-quality sample of cluster members to sources with a PM error $<$20~$\mu$as\,yr$^{-1}$, in order to retain only objects that have the most reliable PM measurements. This leaves us with about 890 stars. For the intrinsic velocity dispersion within central $\sim$35~arcsec ($\sim1.5\times r_{\rm h}$) we found a value of $\sigma_{\mu} = 15.9\pm0.6$~$\mu$as\,yr$^{-1}$. Assuming again a distance of 47.4~kpc to NGC~1850, this translates to $3.57\pm0.13$~km\,s$^{-1}$. We also determined the radial and tangential components ($\sigma_{\rm rad}$ and $\sigma_{\rm tan}$) of the velocity dispersion. Both components show consistent values ($\sigma_{\rm rad} = 15.9\pm0.8$~$\mu$as\,yr$^{-1}$ and $\sigma_{\rm tan} = 15.8\pm0.7$~$\mu$as\,yr$^{-1}$), suggesting that the cluster is isotropic within its central regions.

To construct the radial velocity dispersion profile, we divided the cluster area in four concentric annuli, each containing the same number of stars, and estimated the velocity dispersion within each of these radial bins. The result is illustrated in Fig~\ref{fig:ngc1850_vel_disp_pm} as blue dots. To estimate the central velocity dispersion $\sigma_0$ of the cluster, we opted for fitting the discrete PMs of the stars with a projected \cite{Plummer1911} profile of the following form:
\begin{equation}
\sigma(r)^2 = \frac{\sigma_0^2}{\sqrt{1 + \frac{r^2}{r_{\rm h}^2}}}.
\end{equation}
For the fit, we applied a maximum likelihood approach and set the scale radius $r_{\rm h}$ as well as $\sigma_0$ as free parameters. The fit to our PM data suggest a $\sigma_0 = 16.5\pm0.7$~$\mu$as\,yr$^{-1}$ ($3.71\pm0.16$~km\,s$^{-1}$) and a scale radius $r_{\rm h} = 52.7^{+26.12}_{-18.8}$~arcsec. From our \citet{Plummer1911} profile fit to the stellar density (which is based on stars that cover approximately the same mass range as our kinematic profile), we determined a scale radius of $r_{\rm h}=23.0\pm0.7$~arcsec. This result is in good agreement with the effective (half-light) radius ($r_{\rm e}$) found by \citet{Correnti17}. They fitted a \citet{King62} profile to the stellar radial surface number density profile and reported a value of $r_{\rm e}=20.5\pm1.4$~arcsec. The estimate derived from the PM kinematic analysis is significantly larger and not consistent with the results from the number density profile, which already points to a possible overestimation of the true dispersion, especially within the outer bins of the profile, where the dispersion is smaller.  

To firmly assess the reliability of results for the global and central velocity dispersion, we contrast them with other measurements based on line-of-sight velocities (assuming the system is isotropic within the regions we consider, all components of the velocity dispersion should follow the same profile). 
For the first comparison, we analysed the MUSE data set from \citet{Kamann23} and derived the global velocity dispersion as well as the dispersion profile, employing the same tools as for the PM data. We started by filtering the catalogue for well-measured cluster members. In particular, we kept only stars with (i) an uncertainty in the line-of-sight velocity measurement $<$3~km\,s$^{-1}$, (ii) a line-of-sight velocity variation $<$0.5 and (iii) a $m_{\rm F814W}$ magnitude $<$20~mag. Additionally, we made use of the membership probability calculated by \cite{Kamann23} and selected only stars that have probability to belong to the cluster $>90$\%. Using this sample, we derived a global velocity dispersion within the central $\sim$30~arcsec of $\sigma_{\rm LOS} = 2.1\pm0.2$~km\,s$^{-1}$. The results for the velocity dispersion profile, along with the best-fit Plummer profile is shown in Fig.~\ref{fig:ngc1850_vel_disp_rv}. The MUSE data suggest $\sigma_0 = 2.6^{+0.5}_{-0.3}$~km\,s$^{-1}$ and $r_{\rm h} = 16.3^{+11.9}_{-7.0}$~arcsec. 
This result is consistent with the recent measurement from \cite{Song21}, who found $\sigma_0 = 2.6\pm0.3$~km\,s$^{-1}$. \citet{McLaughlin05} reported for NGC~1850 an observed global intrinsic velocity dispersion of $\sigma_{\rm LOS} =3.00^{+0.70}_{-0.70}$~km\,s$^{-1}$ (measured within an aperture of 40~arcsec) and a central velocity dispersion of $\sigma_0 = 3.19^{+0.77}_{-0.75}$~km\,s$^{-1}$ \citep[see also][]{Fischer93}. 
Based on population synthesis models of the mass-to-light ratio of NGC~1850, \citet{McLaughlin05} predicted a $\sigma_0$ of $\sim$2.2~km\,s$^{-1}$. 

This compilation strongly suggests that the velocity dispersion of NGC~1850 is of the order of $\sim$2.5~km\,s$^{-1}$, corroborating our initial suspicion that the value we calculated using PMs ($3.73\pm0.11$~km\,s$^{-1}$) overestimates the true dispersion of the cluster (to translate our PM-based dispersion to the value of 2.5~km\,s$^{-1}$, a distance to NGC~1850 of only $\sim$31.8~kpc would be required, which is not consistent with the distance to the LMC). For a reliable determination of such small dispersions, an accurate assessment of the measurement uncertainties is crucial, since already small over- or underestimations can significantly affect the final result (which can also be seen in the rather flat profile resulting from the PM data and the associated large value for the scale radius). 

We conclude that the PM catalogues presented in this work do not have the required precision for an accurate evaluation of the internal kinematics of extra-galactic star clusters. However, they provide a promising first step in this direction and with additional epochs of observations this will soon be within reach of HST \citep[or a combination of HST and JWST, e.g.][]{Libralato23, Libralato23b}. This will open up new possibilities for investigations of stellar dynamics within young ($<$1~Gyr) and intermediate-age (few Gyr) massive star clusters within the Magellanic Clouds. This type of clusters (which is either not present in the Milky Way or not easily accessible by observations) is of special relevance since these objects are dynamically young and thus still contain any dynamical signatures imprinted by their formation process.

Finally, based on the results obtained from the MUSE spectroscopic data set, we estimated the total dynamical mass of NGC~1850. Assuming an isotropic Plummer model, the mass is given by \citep[see, e.g.][]{Dejonghe87, Kacharov14}:
\begin{equation}
M = \frac{64r_{\rm h}\sigma_0^2}{3\pi\rm G}.
\end{equation}
We found $\mathrm{log} (M/M_\sun) = 4.67^{+0.40}_{-0.24}$, which is in good agreement with results reported in the literature \citep[e.g.][]{McLaughlin05, Correnti17, Song21, Sollima22}.

\subsection{Radial distribution of the red and blue main sequences}

As a final application, we used the PM-cleaned catalogue of NGC~1850 to study the radial distribution of stars populating the red and blue main sequences. Such split main sequences have been found to be a typical feature within clusters of this age. It has been suggested that the two different branches are formed by stars with different rotation rates, whereas the blue sequence is formed of non- or slowly-rotating stars and the red sequence harbours stars with high rotation rates \citep[see, e.g.][]{D'Antona15, Milone16, Correnti17}. This scenario has later been confirmed by spectroscopic observations \citep[e.g.][]{Marino18, Kamann20, Kamann23}.

Any difference in the spatial distribution of the stars within the two branches can provide insights into the underlying process that creates the range of rotation rates. In the case of NGC~1850, conflicting results regarding the segregation of the two populations exist in the literature. While \citet{Correnti17} did not detect any difference in the radial distribution of the blue and red main sequence stars, \cite{Kamann23} found that the stars on the red branch are more centrally concentrated. 

Here we perform an independent assessment of the radial distributions exploiting our astro-photometric catalogue of NGC~1850. The kinematic information allows us to filter out any unwanted contribution from field stars that might influence the final result. We followed the methods described in \citet{Dresbach2023} and quantify the level of radial segregation by means of the so-called $A^+$ parameter \citep[which was initially defined to study blue straggler stars, see, e.g.][]{Alessandrini16, Ferraro18, Dresbach2022}. In our case, $A^+$ can be written as:
\begin{equation}
    A^{+} = \int_{x_{\rm min}}^{x_{\rm max}} \Big( \phi_{\rm red}(x') - \phi_{\rm blue}(x') \Big)\,{\rm d}x',
\end{equation}
where $x'$ refers to the logarithmic distance from the cluster centre, normalised to the scale radius (log$(r/r_{\rm h})$), and $\phi_{\rm red}$ and $\phi_{\rm blue}$ denote the cumulative radial distribution of the red and blue main sequence stars, respectively. 
Thus, $A^+$ gives the area lying between these two curves.

\begin{figure}[h]
\centering
\includegraphics[width=\columnwidth]{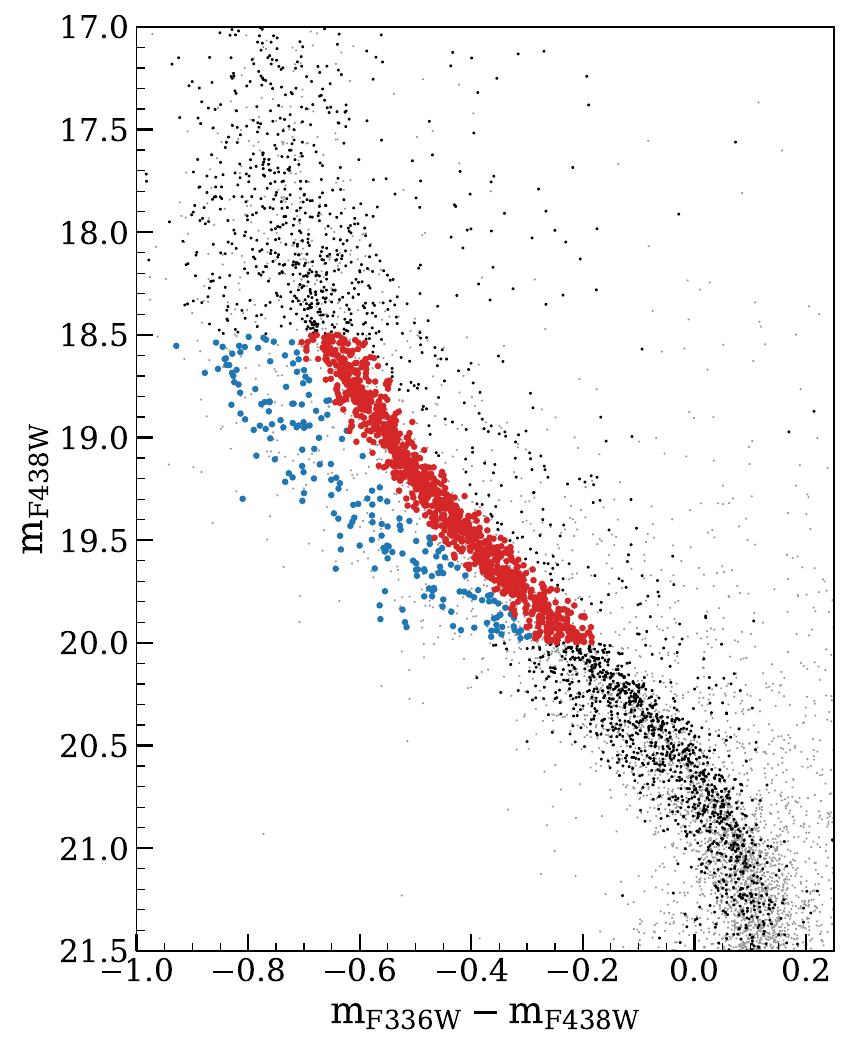}
\caption{$m_{\rm F438W}$ vs $m_{\rm F336W} - m_{\rm F438W}$ CMD of NGC~1850, zoomed in the region where the split of the main sequence is most evident. The selected stars belonging to the red (blue) branch are indicated by red (blue) dots.\label{fig:ngc1850_MS}}
\end{figure}

\begin{figure}[h]
\centering
\includegraphics[width=\columnwidth]{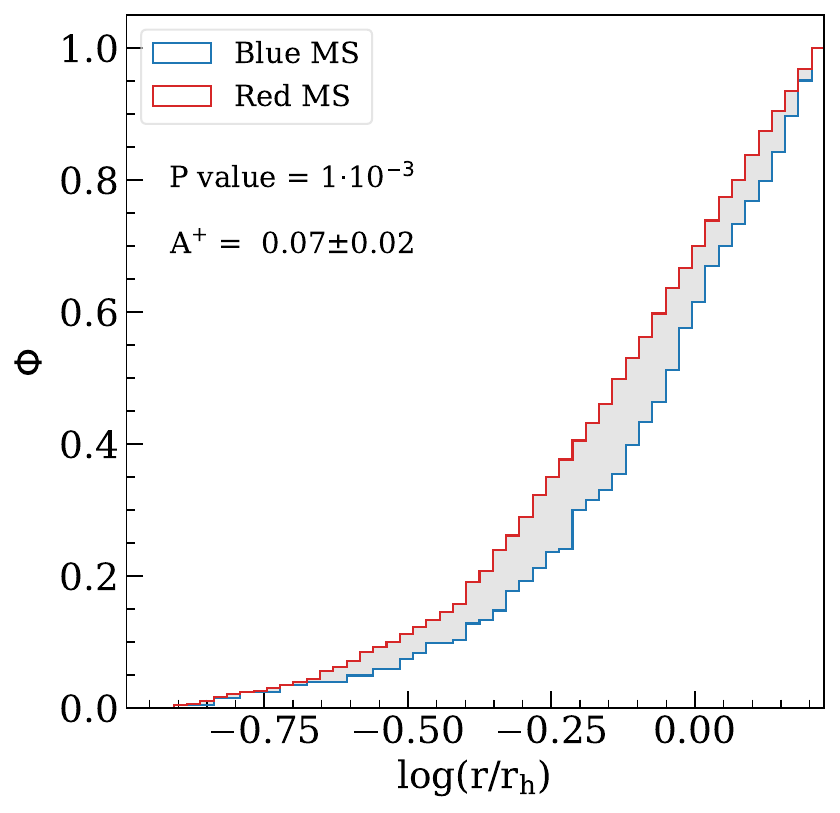}
\caption{Cumulative radial distribution of the blue and red main sequence stars. Also indicated in the figure is the probability that the two samples are drawn from the same underlying population (P value), and the value of the $A^+$ parameter (which corresponds to the grey-shaded region). \label{fig:ngc1850_blue_red_MS}}
\end{figure}

We identified stars belonging to the red and blue sequences as follows: First, we only considered stars in the magnitude interval 18.5~mag$\le m_{\rm F438W} \le$20.0~mag, where the splitting is most evident (see Fig.~\ref{fig:ngc1850_MS}). Then, we constructed a fiducial line of the red main sequence. For this, we selected (by hand) a preliminary sample of stars belonging to that sequence and determined its median $m_{\rm F336W} - m_{\rm F438W}$ colour and $m_{\rm F438W}$ magnitude in bins of 0.2~mag. Interpolating the median points with a cubic spline yields the fiducial line. We verticalized the CMD by calculating the distance $\Delta_{\rm col}$ of each star from the fiducial line and fitted the $\Delta_{\rm col}$ distribution with a three-component (one for the blue and red branches as well as for the binary sequence) Gaussian-mixture-model. From the main Gaussian component (which corresponds to the red sequence), we defined a red and a blue boundary, corresponding to $\pm2.5$ times the standard deviation of that Gaussian. We used the blue boundary as the division line to separate blue and red main sequence stars and selected stars lying between the two boundaries as red main sequence stars (see Fig.~\ref{fig:ngc1850_MS}). With this selection, we constructed the cumulative radial distributions of the two sequences and calculated the corresponding $A^+$ parameter. This is illustrated in Fig.~\ref{fig:ngc1850_blue_red_MS}. As we can see from the figure, 
the curve corresponding to the blue main sequence traces a flatter distribution. The resulting $A^+$ parameter is $0.07\pm0.02$. We estimated the uncertainty associated to the value of $A^+$ from a series of bootstrap realisations of the two samples. From a two-sided Kolmogorov–Smirnov test we obtained a probability of $\sim1\times10^{-3}$ that the two distributions are drawn from the same underlying population. 
This indicates that stars populating the red sequence (fast-rotating stars) are more centrally concentrated compared to the stars along the blue sequence (slow-rotating stars). We also verified that our obtained result does not depend on the astrometric and photometric quality selections that we applied to define our high-quality sample. For this, we repeated the above analysis using our entire catalogue and only applied cuts based on the relative PM ($\le80~\mu$as\,yr$^{-1}$) and distance from the cluster centre ($\le1.5\times r_{\rm h}$) to select likely cluster members. Here, we found $A^+ = 0.06\pm0.02$ and a probability of $\sim2\times10^{-3}$ that both samples are drawn from the same distribution. 

Using the PM measurements, we also calculated the velocity dispersions of the two samples. For stars along the blue main sequence, we found $\sigma  = 15.8\pm2.2$~$\mu$as\,yr$^{-1}$ and for stars populating the red sequence, we found $\sigma  = 14.4\pm1.0$~$\mu$as\,yr$^{-1}$. Both values are consistent with each other within the uncertainties. Thus, we are not able to assess from our data whether there is any difference in the kinematics between the stars in the two sequences.

Our results regarding the radial segregation corroborate the findings from \citet{Kamann23}. However, looking at similar studies that have investigated other young LMC clusters there is no consistent picture regarding any spacial segregation of stars populating the two different branches of the split main sequence. On the one hand, \citet{Li17} did not find any difference in the radial distributions of the two populations within NGC~1856. Similarly, \citet{Dresbach2023} did not find any evidence for spatial segregation in NGC~419. On the other hand, \citet{Milone17} reported for NGC~1866 an increasing fraction of blue-to-red main sequence stars with increasing distance from the cluster centre, corresponding to a higher concentration of rapidly rotating stars in the central parts of the cluster. The authors further found evidence that the distribution of binary stars follows the one from the blue main sequence stars. The indication that the population of slow-rotating stars could be connected to the population of binaries would support the scenario where fast-rotating stars are slowed-down by tidal torques in binary systems, creating the population of slow rotating stars \citep[see, e.g.][]{D'Antona15, D'Antona17}. However, \citet{Kamann21} derived the binary fractions within the blue and red main sequences of NGC~1850 and found similar fractions, which contradicts the binary-driven origin of the slow-rotating stars.


\section{Summary and conclusions}\label{sec:conclusions}

In this paper, we presented PM calculations of a sample of 23 star clusters in the LMC, using multi-epoch HST data. For 19 clusters, one epoch of observations comes from our dedicated GO-16478 programme. We complement them with four more clusters that have already-existing observations in the HST archive suitable for PM calculations. Since the PMs result from a combination of a heterogeneous set of data, their final quality differs from cluster to cluster. The data that we have at our disposal span time-baselines between $\sim$4.7~yr and $\sim$18.2~yr, depending on the cluster. Employing state-of-the-art photometric and astrometric reduction techniques, we achieve mean nominal PM precisions of $\sim$55~$\mu$as\,yr$^{-1}$ down to $\sim$11~$\mu$as\,yr$^{-1}$, for the best-measured stars within each of our targets. We determined relative PMs for a total of $\sim$855\,000 sources, distributed among the 23 clusters within our sample, whereas the field-of-view of NGC~2209 contains the least number of stars ($\sim$7\,000) and the one of NGC~2005 the most ($\sim$112\,000). 
With this study, we make the astro-photometric catalogues publicly available. 
We also determined the centres and structural parameters of the clusters within our sample by fitting a Plummer profile to their stellar density profiles. 

We showcased the capabilities and limitations of our PM catalogues and presented a selection of science applications using NGC~1850 as an example. 
Thanks to the precision of the PM measurements, we were able to disentangle the kinematics of the various stellar populations that are within the field of NGC~1850. We separated the motions of the cluster members from the ones of field stars, as well as the motions of stars belonging to a close group of very young stars, dubbed NGC~1850B. Combining the PMs with MUSE line-of-sight velocities, we showed that the two clusters show very different tangential motions within the LMC disc, thus suggesting they are not dynamically related. We determined an absolute motion of NGC~1850 of $\mu_{\alpha}\rm cos\delta = 2.011\pm0.028$~mas\,yr$^{-1}$, $\mu_{\delta} = 0.119\pm0.030$~mas\,yr$^{-1}$. The motion measured for the field stars corresponds to a centre-of-mass motion of the LMC of $\mu_{\alpha}\rm cos\delta = 1.871\pm0.028\pm0.006$~mas\,yr$^{-1}$, $\mu_{\delta} = 0.291\pm0.030\pm0.006$~mas\,yr$^{-1}$. We further determined the velocity dispersion of the field stars and obtained $0.128\pm0.003$~mas\,yr$^{-1}$ ($30.3\pm0.7$~km\,s$^{-1}$).

Based on the PM measurements of NGC~1850, we further presented an attempt to determine the velocity dispersion within the plane of the sky of an extra-galactic star cluster. We found for the central velocity dispersion a value of $16.5\pm0.7$~$\mu$as\,yr$^{-1}$ ($3.71\pm0.16$~km\,s$^{-1}$), most probably overestimating the true dispersion. 
From the MUSE line-of-sight velocity dispersion, we estimated a dynamical mass of NGC~1850 of ${\rm log}(M/M_{\sun})=4.67^{+0.40}_{-0.24}$.

Finally, we utilised the PM-cleaned catalogue of NGC~1850 member stars to investigate the radial distribution of stars that populate the red and blue branches of the cluster's main sequence. By means of the $A^+$ parameter we quantised any spatial segregation and found that stars on the red main sequence are more centrally concentrated than stars on the blue sequence.

Precise measurements of stellar PMs towards star clusters within the Magellanic Clouds are now feasible with HST. This enables studies of the resolved dynamics of the Magellanic Cloud star-cluster system using PMs. Especially, PMs can be used to obtain clean samples of cluster stars to investigate peculiar features in their CMDs, and also to study the kinematic pattern of the Magellanic Clouds as traced by star clusters
to gain a better understanding of their dynamical history and evolution.
Additional epochs of observations will even pave the road to study internal kinematics of star clusters at the distance of the Magellanic Clouds.
In future studies, we will combine the measured PMs with line-of-sight velocities to investigate the full 3D kinematic structure of the star clusters within the LMC.


\begin{acknowledgements}  
We thank the anonymous referee for constructive comments and suggestions that improved the quality of our paper.
This research was funded by DLR grant 50 OR 2216.
Support for this work was provided by NASA through grants for program GO-16478 from the Space Telescope Science Institute (STScI), which is operated by the Association of Universities for Research in Astronomy (AURA), Inc., under NASA contract NAS5-26555. This study was supported by the Klaus Tschira Foundation. SK acknowledges funding from UKRI in the form of a Future Leaders Fellowship (grant no. MR/T022868/1). SM was supported by a Gliese Fellowship at the Zentrum für Astronomie, University of Heidelberg, Germany. This work is based on observations made with the NASA/ESA \textit{Hubble} Space Telescope, obtained from the Data Archive at the Space Telescope Science Institute. This work has made use of data from the European Space Agency (ESA) mission \textit{Gaia} (\url{https://www.cosmos.esa.int/gaia}), processed by the Gaia Data Processing and Analysis Consortium (DPAC, \url{https://www.cosmos.esa.int/web/gaia/dpac/consortium}). Funding for the DPAC has been provided by national institutions, in particular the institutions participating in the Gaia Multilateral Agreement.
This research made use of \texttt{astropy},\footnote{\href{http://www.astropy.org}{http://www.astropy.org}} a community-developed core \texttt{python} package for Astronomy \citep{Astropy13, Astropy18}, \texttt{iphython} \citep{Perez07}, \texttt{Jupyter Notebook} \citep{Kluyver16}, \texttt{matplotlib} \citep{Hunter07}, \texttt{numpy} \citep{Harris2020} and \texttt{scipy} \citep{Virtanen20}.
\end{acknowledgements}



\bibliographystyle{aa}
\bibliography{references}

\begin{appendix} 

\section{Description of the astrometric and photometric catalogues\label{app:catalogues}}

We release the astro-photometric catalogues of the 23 LMC clusters assembled in this publication to the scientific community. 
All data products are available as a High Level Science Product at MAST under the following DOI: \href{https://doi.org/10.17909/tx9s-7540}{10.17909/tx9s-7540} \footnote{see also: \href{https://archive.stsci.edu/hlsp/hamsters}{https://archive.stsci.edu/hlsp/hamsters}}. For each cluster, the catalogue is split into an astrometric (PM) file and one individual photometric file for each combination of filter and epoch. The catalogues for a given cluster are all sorted in the same order such that the same line in the different catalogues corresponds to the same source. 

The content of the PM catalogue is shown in Table~\ref{tab:PMcat}. The X and Y coordinates give the positions of the sources on the master frame that has a pixel scale of 40~mas\,pixel$^{-1}$ and the cluster centre placed approximately at (10\,000, 10\,000). The X,Y coordinates refer to positions at the reference epoch we adopted for the PM calculations, which is stated in the header of the file. The RA and Dec coordinates are given in the ICRS and the \textit{Gaia} DR3 reference epoch (2016.0). The astrometric catalogue further contains both the raw and a posteriori corrected PMs along with their associated uncertainties. A ``Correction Flag'' indicates whether the PM has been a posteriori corrected for small-scale systematic effects (flag = 1) or not (flag = 0). All PMs in the catalogues are given as relative motions. They can be converted to absolute values applying the PM ZPs given in the header of the file. Also included in the astrometric file are a number of quality parameters, such as the reduced $\chi^2$ of the PM fits. The quantities N$_\mathrm{f}$ and N$_\mathrm{u}$ refer to the number of individual measurements initially available for the PM fit and the final number of data points actually used to determine the PM, respectively. The time baseline for the PM determination is given by $\Delta$time. Finally, the ID number is the same for a given source among all catalogues of an individual cluster. 

Table \ref{tab:Photcat} describes the columns included in the photometric file. The catalogues contain the VEGA-calibrated magnitudes for each source with measured PM along with photometric quality parameters. The VEGA-mag ZP used to transform the instrumental magnitudes to the VEGA system is given in the header of the catalogue. The file lists the photometric \texttt{RMS}, as well as the quality of fit \texttt{QFIT} and the radial excess \texttt{RADXS} parameters. \texttt{QFIT} indicates how well a source is fit by the PSF model and can have values between 0.0 and 1.0, where a value of 1.0 means a perfect fit. \texttt{RADXS} provides a measure of how extended a source is with respect to the model PSF. The proximity parameter is denoted by the quantity $o$. It states the fraction of flux within the PSF fitting radius that is due to neighbouring sources (before neighbour subtraction). N$_\mathrm{f}$ and N$_\mathrm{u}$ correspond to the number of exposures a source was found in and the number of exposures used for the measurement of the flux, respectively. The photometric catalogues also contain a saturation flag that works according to the following scheme: a flag of 1 means the star is unsaturated in all exposures, the highest number present means the star is saturated. If there are more exposure-time groups in the data set, the flag increases by 1 for each shorter exposure-time group the star becomes saturated. For example, if there are long, medium and short exposures, the flags are: 1 = unsaturated in the long exposures, 2 = unsaturated in the medium exposures, 3 = unsaturated in the short exposures, 4 = saturated. Since \texttt{KS2} does not measure saturated stars, magnitudes for these sources come from the first-pass photometric run. 
If a star has no measurement in a specific filter and epoch, all the above quantities are set to 0.0.

\begin{table*}\small
\centering
\caption{Description of the content of the PM catalogues\label{tab:PMcat}}
\begin{tabular}{lccr}
\hline\hline
\noalign{\smallskip}
Column & Name& ~~~~~~~~~~Unit~~~~~~~~~ & ~~~~~~~~~~~~~~~~~~~~~~~~~~~~~~~~~~~Description \\
\noalign{\smallskip}
\hline
\noalign{\smallskip}
1 & RA & [deg] & Right ascension (ICRS) at epoch 2016.0\\
2 & Dec & [deg] & Declination (ICRS) at epoch 2016.0\\
3 & X & [pixel] & X position on the master frame;\\
& & & pixel scale of 40~mas\,pixel$^{-1}$ and the cluster centre at $\sim$(10\,000, 10\,000)\\
4 & Y & [pixel] & Y position on the master frame;\\
& & & pixel scale of 40~mas\,pixel$^{-1}$ and the cluster centre at $\sim$(10\,000, 10\,000)\\
5 & ($\mu_{\alpha}$\,cos$\delta$)$_{\mathrm{corr}}$ & [mas\,yr$^{-1}$] & Corrected PM along RA direction \\
6 & ($\sigma_{\mu_{\alpha}\,\mathrm{cos}\delta}$)$_{\mathrm{corr}}$ & [mas\,yr$^{-1}$] & Error of corrected PM along RA direction \\
7 & ($\mu_{\delta}$)$_{\mathrm{corr}}$ & [mas\,yr$^{-1}$] & Corrected PM along Dec direction \\
8 & ($\sigma_{\mu_{\delta}}$)$_{\mathrm{corr}}$ & [mas\,yr$^{-1}$] & Error of corrected PM along Dec direction\\
9 & $\chi^2_{\mu_{\alpha}\,\mathrm{cos}\delta}$ & & Reduced $\chi^2$ of the PM fit along RA direction\\
10 & $\chi^2_{\mu_{\delta}}$ & & Reduced $\chi^2$ of the PM fit along Dec direction\\
11 & N$_{\mathrm{f}}$ & & Initial number of exposures available for PM fit \\
12 & N$_{\mathrm{u}}$ & & Final number of exposures actually used for PM fit \\
13 & $\Delta$time& & Time baseline for PM fit\\
14 & ($\mu_{\alpha}$\,cos$\delta$)$_{\mathrm{raw}}$& [mas\,yr$^{-1}$] & Raw PM along RA direction\\
15 & ($\sigma_{\mu_{\alpha}\,\mathrm{cos}\delta}$)$_{\mathrm{raw}}$ & [mas\,yr$^{-1}$] & Error of raw PM along RA direction\\
16 & ($\mu_{\delta}$)$_{\mathrm{raw}}$ & [mas\,yr$^{-1}$] & Raw PM along Dec direction \\
17 & ($\sigma_{\mu_{\delta}}$)$_{\mathrm{raw}}$ & [mas\,yr$^{-1}$] & Error of raw PM along Dec direction \\
18 & Correction Flag & & 1 = PM has been a posteriori corrected, 0 = otherwise \\
19 & ID & & Identification number\\
\noalign{\smallskip}
\hline

\end{tabular}

\end{table*}

\begin{table*}\small
\centering
\caption{Description of the content of the photometric catalogues \label{tab:Photcat}}
\begin{tabular}{lcr}
\hline\hline
\noalign{\smallskip}
Column & ~~~~~~~~~Name~~~~~~~~~ & ~~~~~~~~~~~~~~~~~~~~~~~~~~~~~~~~~~~Description \\
\noalign{\smallskip}
\hline
\noalign{\smallskip}
1 & $m$ &  Calibrated VEGA magnitude\\
2 & \texttt{RMS}$_m$ & Photometric RMS\\
3 & \texttt{QFIT}$_m$ &  Quality-of-PSF-fit parameter\\
4 & \texttt{RADXS}$_m$&  Radial excess parameter\\
5 & $o$&  Proximity parameter; \\
 & & fraction of flux within the PSF aperture due to neighbours\\
6 & N$_\mathrm{f}$ & Number of exposures a source was found in\\
7 & N$_\mathrm{u}$ & Number of exposures used to measure flux  \\
8 & \texttt{SAT}& Saturation flag \\
9 & ID & Identification number \\
\noalign{\smallskip}
\hline

\end{tabular}

\end{table*}

\section{List of observations\label{app:obs}}
Tables \ref{tab:hodge11obs}--\ref{tab:ngc2257obs} provide, for each cluster, the full list of exposures used for the PM determinations. The tables list the programme ID number, the PI, the epoch of observations, the used instrument, camera and filter combination, as well as the number of exposures along with the respective exposure times.

\begin{table*}
\centering
\caption{Observations of HODGE 11\label{tab:hodge11obs}}
\begin{tabular}{@{}l@{ }c@{ }c@{ }c@{ }c@{ }c@{ }c@{ }}
\hline\hline
\noalign{\smallskip}
Programme ID & ~~~~~~~~PI~~~~~~~~ & ~~~~~~~Epoch~~~~~~~ & ~~~~~Instrument/Camera~~~~~ & ~~~~~Filter~~~~~ & ~~~~~Exposures~~~~~\\
& & (yyyy/mm) & & & (N\,$\times$\,t$_{\mathrm{exp}}$) \\
\noalign{\smallskip}
\hline
\noalign{\smallskip}
GO-14164 &  A. Sarajedini & 2016/12 & WFC3/UVIS & F336W & 3\,$\times$\,700\,s \\
&   &  & &  & 12\,$\times$\,729\,s \\
&  & 2016/06 & ACS/WFC & F606W & 2\,$\times$\,50\,s \\
&  &  &  & & 6\,$\times$\,345\,s \\
&  &  &  & & 6\,$\times$\,370\,s \\
&  & 2016/07 & ACS/WFC & F814W & 2\,$\times$\,70\,s \\
&  &  &  &  & 6\,$\times$\,345\,s \\
&  &  &  &  & 6\,$\times$\,377\,s \\
&  &  &  &  & 6\,$\times$\,410\,s \\
 \noalign{\smallskip}
\hline
\noalign{\smallskip}
GO-16748 & F. Niederhofer & 2022/03 & WFC3/UVIS & F814W & 2\,$\times$\,40\,s\\
&  &  &  & & 3\,$\times$\,446\,s\\
&  &  &  & & 2\,$\times$\,447\,s\\
\noalign{\smallskip}
\hline

\end{tabular}

\end{table*}


\begin{table*}
\centering
\caption{Observations of NGC 1651\label{tab:ngc1651obs}}
\begin{tabular}{@{}l@{ }c@{ }c@{ }c@{ }c@{ }c@{ }c@{ }}
\hline\hline
\noalign{\smallskip}
Programme ID & ~~~~~~~~PI~~~~~~~~ & ~~~~~~~Epoch~~~~~~~ & ~~~~~Instrument/Camera~~~~~ & ~~~~~Filter~~~~~ & ~~~~~Exposures~~~~~\\
& & (yyyy/mm) & & & (N\,$\times$\,t$_{\mathrm{exp}}$) \\
\noalign{\smallskip}
\hline
\noalign{\smallskip}
GO-12257 & L. Girardi  & 2011/10 & WFC3/UVIS & F475W & 1\,$\times$\,120\,s \\
 &   &  &  &  & 1\,$\times$\,600\,s \\
 &   &  &  &  & 1\,$\times$\,720\,s \\
 &   &  &  & F814W & 1\,$\times$\,30\,s \\
  &   &  &  & & 2\,$\times$\,700\,s \\
 \noalign{\smallskip}
\hline
\noalign{\smallskip}
GO-16748 & F. Niederhofer & 2022/01 & WFC3/UVIS & F814W & 2\,$\times$\,40\,s\\
&  &  &  & & 4\,$\times$\,454\,s\\
&  &  &  & & 1\,$\times$\,455\,s\\
\noalign{\smallskip}
\hline

\end{tabular}

\end{table*}


\begin{table*}
\centering
\caption{Observations of NGC 1718\label{tab:ngc1718obs}}
\begin{tabular}{@{}l@{ }c@{ }c@{ }c@{ }c@{ }c@{ }c@{ }}
\hline\hline
\noalign{\smallskip}
Programme ID & ~~~~~~~~PI~~~~~~~~ & ~~~~~~~Epoch~~~~~~~ & ~~~~~Instrument/Camera~~~~~ & ~~~~~Filter~~~~~ & ~~~~~Exposures~~~~~\\
& & (yyyy/mm) & & & (N\,$\times$\,t$_{\mathrm{exp}}$) \\
\noalign{\smallskip}
\hline
\noalign{\smallskip}
GO-12257 & L. Girardi  & 2011/12 & WFC3/UVIS & F475W & 1\,$\times$\,120\,s \\
 &   &  &  &  & 1\,$\times$\,600\,s \\
 &   &  &  &  & 1\,$\times$\,720\,s \\
 &   &  &  & F814W & 1\,$\times$\,30\,s \\
  &   &  &  & & 2\,$\times$\,700\,s \\
 \noalign{\smallskip}
\hline
\noalign{\smallskip}
GO-16748 & F. Niederhofer & 2021/09 & WFC3/UVIS & F814W & 2\,$\times$\,40\,s\\
&  &  &  & & 3\,$\times$\,446\,s\\
&  &  &  & & 2\,$\times$\,447\,s\\
\noalign{\smallskip}
\hline

\end{tabular}

\end{table*}


\begin{table*}
\centering
\caption{Observations of NGC 1783\label{tab:ngc1783obs}}
\begin{tabular}{@{}l@{ }c@{ }c@{ }c@{ }c@{ }c@{ }c@{ }}
\hline\hline
\noalign{\smallskip}
Programme ID & ~~~~~~~~PI~~~~~~~~ & ~~~~~~~Epoch~~~~~~~ & ~~~~~Instrument/Camera~~~~~ & ~~~~~Filter~~~~~ & ~~~~~Exposures~~~~~\\
& & (yyyy/mm) & & & (N\,$\times$\,t$_{\mathrm{exp}}$) \\
\noalign{\smallskip}
\hline
\noalign{\smallskip}
GO-9891 & G. Gilmore  & 2003/10 & ACS/WFC & F555W & 1\,$\times$\,250\,s \\
 &   &  &  & F814W & 1\,$\times$\,170\,s \\
\noalign{\smallskip}
\hline
\noalign{\smallskip}
GO-10595 & P. Goudfrooij  & 2006/01 & ACS/WFC & F435W & 1\,$\times$\,90\,s \\
 &   &  &  &  & 2\,$\times$\,340\,s \\
 &   &  &  & F555W & 1\,$\times$\,40\,s \\
  &   &  &  &  & 2\,$\times$\,340\,s \\
 &   &  &  & F814W & 1\,$\times$\,8\,s \\
  &   &  &  &  & 2\,$\times$\,340\,s \\
\noalign{\smallskip}
\hline
\noalign{\smallskip}
GO-12257 & L. Girardi  & 2011/10 & WFC3/UVIS & F336W & 2\,$\times$\,1190\,s \\
 &   &  &  & & 1\,$\times$\,1200\,s \\
\noalign{\smallskip}
\hline
\noalign{\smallskip}
GO-16255 & E. Dalessandro  & 2021/01 & WFC3/UVIS & F438W & 6\,$\times$\,938\,s \\
\noalign{\smallskip}
\hline

\end{tabular}

\end{table*}


\begin{table*}
\centering
\caption{Observations of NGC 1805\label{tab:ngc1805obs}}
\begin{tabular}{@{}l@{ }c@{ }c@{ }c@{ }c@{ }c@{ }c@{ }}
\hline\hline
\noalign{\smallskip}
Programme ID & ~~~~~~~~PI~~~~~~~~ & ~~~~~~~Epoch~~~~~~~ & ~~~~~Instrument/Camera~~~~~ & ~~~~~Filter~~~~~ & ~~~~~Exposures~~~~~\\
& & (yyyy/mm) & & & (N\,$\times$\,t$_{\mathrm{exp}}$) \\
\noalign{\smallskip}
\hline
\noalign{\smallskip}
GO-13727 & J. Kalirai  & 2015/10 & WFC3/UVIS & F336W & 1\,$\times$\,100\,s \\
 &   &  &  &  & 3\,$\times$\,947\,s \\
 \noalign{\smallskip}
\hline
\noalign{\smallskip}
GO-14710 & A. Milone  & 2017/10 & WFC3/UVIS & F814W & 1\,$\times$\,90\,s \\
 &   &  &  &  & 1\,$\times$\,666\,s \\
 \noalign{\smallskip}
\hline
\noalign{\smallskip}
GO-16748 & F. Niederhofer & 2022/01 & WFC3/UVIS & F814W & 2\,$\times$\,40\,s\\
&  &  &  & & 3\,$\times$\,446\,s\\
&  &  &  & & 2\,$\times$\,447\,s\\
\noalign{\smallskip}
\hline

\end{tabular}

\end{table*}


\begin{table*}
\centering
\caption{Observations of NGC 1806\label{tab:ngc1806obs}}
\begin{tabular}{@{}l@{ }c@{ }c@{ }c@{ }c@{ }c@{ }c@{ }}
\hline\hline
\noalign{\smallskip}
Programme ID & ~~~~~~~~PI~~~~~~~~ & ~~~~~~~Epoch~~~~~~~ & ~~~~~Instrument/Camera~~~~~ & ~~~~~Filter~~~~~ & ~~~~~Exposures~~~~~\\
& & (yyyy/mm) & & & (N\,$\times$\,t$_{\mathrm{exp}}$) \\
\noalign{\smallskip}
\hline
\noalign{\smallskip}
GO-9891 & G. Gilmore  & 2003/08 & ACS/WFC & F555W & 1\,$\times$\,300\,s \\
 &   &  &  & F814W & 1\,$\times$\,200\,s \\
 \noalign{\smallskip}
\hline
\noalign{\smallskip}
GO-10595 & P. Goudfrooij  & 2005/09 & ACS/WFC & F435W & 1\,$\times$\,90\,s \\
 &   &  &  &  & 2\,$\times$\,340\,s \\
 &   &  &  & F555W & 1\,$\times$\,40\,s \\
  &   &  &  &  & 2\,$\times$\,340\,s \\
 &   &  &  & F814W & 1\,$\times$\,8\,s \\
  &   &  &  &  & 2\,$\times$\,340\,s \\
 \noalign{\smallskip}
\hline
\noalign{\smallskip}
GO-12257 & L. Girardi  & 2011/10 & WFC3/UVIS & F336W & 2\,$\times$\,1190\,s \\
 &   &  &  & & 1\,$\times$\,1200\,s \\
\noalign{\smallskip}
\hline

\end{tabular}

\end{table*}


\begin{table*}
\centering
\caption{Observations of NGC 1831\label{tab:ngc1831obs}}
\begin{tabular}{@{}l@{ }c@{ }c@{ }c@{ }c@{ }c@{ }c@{ }}
\hline\hline
\noalign{\smallskip}
Programme ID & ~~~~~~~~PI~~~~~~~~ & ~~~~~~~Epoch~~~~~~~ & ~~~~~Instrument/Camera~~~~~ & ~~~~~Filter~~~~~ & ~~~~~Exposures~~~~~\\
& & (yyyy/mm) & & & (N\,$\times$\,t$_{\mathrm{exp}}$) \\
\noalign{\smallskip}
\hline
\noalign{\smallskip}
GO-14688 & P. Goudfrooij  & 2016/11 & WFC3/UVIS & F336W & 2\,$\times$\,975\,s \\
 &   &  &  &  & 2\,$\times$\,1115\,s \\
 &   &  &  & F814W & 1\,$\times$\,100\,s \\
&   &  &  &  & 1\,$\times$\,660\,s \\
&   &  &  &  & 1\,$\times$\,720\,s \\
 \noalign{\smallskip}
\hline
\noalign{\smallskip}
GO-16748 & F. Niederhofer & 2022/09 & WFC3/UVIS & F814W & 2\,$\times$\,40\,s\\
&  &  &  & & 3\,$\times$\,431\,s\\
&  &  &  & & 2\,$\times$\,432\,s\\
\noalign{\smallskip}
\hline

\end{tabular}

\end{table*}


\begin{table*}
\centering
\caption{Observations of NGC 1841\label{tab:ngc1841obs}}
\begin{tabular}{@{}l@{ }c@{ }c@{ }c@{ }c@{ }c@{ }c@{ }}
\hline\hline
\noalign{\smallskip}
Programme ID & ~~~~~~~~PI~~~~~~~~ & ~~~~~~~Epoch~~~~~~~ & ~~~~~Instrument/Camera~~~~~ & ~~~~~Filter~~~~~ & ~~~~~Exposures~~~~~\\
& & (yyyy/mm) & & & (N\,$\times$\,t$_{\mathrm{exp}}$) \\
\noalign{\smallskip}
\hline
\noalign{\smallskip}
GO-14164 & A. Sarajedini & 2015/12 & ACS/WFC & F606W & 2\,$\times$\,50\,s \\
&  &  &  & & 12\,$\times$\,353\,s \\
&  & 2015/12 & ACS/WFC & F814W & 2\,$\times$\,70\,s \\
&  &  &  &  & 6\,$\times$\,352\,s \\
&  &  &  &  & 6\,$\times$\,385\,s \\
&  &  &  &  & 6\,$\times$\,420\,s \\
\noalign{\smallskip}
\hline
\noalign{\smallskip}
GO-14164 &  A. Sarajedini & 2016/06 & WFC3/UVIS & F336W & 4\,$\times$\,700\,s \\
&   &  & &  & 12\,$\times$\,739\,s \\
 \noalign{\smallskip}
\hline
\noalign{\smallskip}
GO-16748 & F. Niederhofer & 2022/01 & WFC3/UVIS & F814W & 2\,$\times$\,40\,s\\
&  &  &  & & 5\,$\times$\,456\,s\\
\noalign{\smallskip}
\hline

\end{tabular}

\end{table*}


\begin{table*}
\centering
\caption{Observations of NGC 1846\label{tab:ngc1846obs}}
\begin{tabular}{@{}l@{ }c@{ }c@{ }c@{ }c@{ }c@{ }c@{ }}
\hline\hline
\noalign{\smallskip}
Programme ID & ~~~~~~~~PI~~~~~~~~ & ~~~~~~~Epoch~~~~~~~ & ~~~~~Instrument/Camera~~~~~ & ~~~~~Filter~~~~~ & ~~~~~Exposures~~~~~\\
& & (yyyy/mm) & & & (N\,$\times$\,t$_{\mathrm{exp}}$) \\
\noalign{\smallskip}
\hline
\noalign{\smallskip}
GO-9891 & G. Gilmore  & 2003/10 & ACS/WFC & F555W & 1\,$\times$\,300\,s \\
 &   &  &  & F814W & 1\,$\times$\,200\,s \\
 \noalign{\smallskip}
\hline
\noalign{\smallskip}
GO-10595 & P. Goudfrooij  & 2006/01 & ACS/WFC & F435W & 1\,$\times$\,90\,s \\
 &   &  &  &  & 2\,$\times$\,340\,s \\
 &   &  &  & F555W & 1\,$\times$\,40\,s \\
  &   &  &  &  & 2\,$\times$\,340\,s \\
 &   &  &  & F814W & 1\,$\times$\,8\,s \\
  &   &  &  &  & 2\,$\times$\,340\,s \\
 \noalign{\smallskip}
\hline
\noalign{\smallskip}
GO-12257 & L. Girardi  & 2011/04 & WFC3/UVIS & F336W & 1\,$\times$\,900\,s \\
 &   &  &  & & 8\,$\times$\,1032\,s \\
\noalign{\smallskip}
\hline

\end{tabular}

\end{table*}


\begin{table*}
\centering
\caption{Observations of NGC 1850\label{tab:ngc1850obs}}
\begin{tabular}{@{}l@{ }c@{ }c@{ }c@{ }c@{ }c@{ }c@{ }}
\hline\hline
\noalign{\smallskip}
Programme ID & ~~~~~~~~PI~~~~~~~~ & ~~~~~~~Epoch~~~~~~~ & ~~~~~Instrument/Camera~~~~~ & ~~~~~Filter~~~~~ & ~~~~~Exposures~~~~~\\
& & (yyyy/mm) & & & (N\,$\times$\,t$_{\mathrm{exp}}$) \\
\noalign{\smallskip}
\hline
\noalign{\smallskip}
GO-11925 &  S. Deustua & 2009/10 & WFC3/UVIS & F467M & 4\,$\times$\,10\,s \\
& &  &  &  & 2\,$\times$\,20\,s \\
& &  &  &  & 2\,$\times$\,50\,s \\
& &  &  &  & 2\,$\times$\,100\,s \\
& &  &  &  & 2\,$\times$\,300\,s\\
& &  &  &  & 2\,$\times$\,500\,s\\
\noalign{\smallskip}
\hline
\noalign{\smallskip}
GO-13282 & Y.-H. Chu & 2014/06 & WFC3/UVIS & F555W & 1\,$\times$\,20\,s\\
 &  &  &  & & 2\,$\times$\,350\,s\\
  &  &  &  & & 1\,$\times$\,417\,s\\
 &  &  &  & F814W & 1\,$\times$\,20\,s\\
  &  &  &  &  & 3\,$\times$\,350\,s\\
\noalign{\smallskip}
\hline
\noalign{\smallskip}
GO-14069 & N. Bastian & 2015/12 & WFC3/UVIS & F336W & 1\,$\times$\,260\,s\\
 &  &  &  &  & 1\,$\times$\,370\,s\\
 &  &  &  &  & 1\,$\times$\,600\,s\\
 &  &  &  &  & 1\,$\times$\,650\,s\\
 &  &  &  &  & 1\,$\times$\,675\,s\\

 &  &  & & F438W & 1\,$\times$\,45\,s\\
 &  &  &  &  & 1\,$\times$\,90\,s\\
 &  &  &  &  & 1\,$\times$\,110\,s\\
 &  &  &  &  & 2\,$\times$\,400\,s\\
\noalign{\smallskip}
\hline
\noalign{\smallskip}
GO-14174 & P. Goudfrooij & 2015/10 & WFC3/UVIS & F814W & 1\,$\times$\,7\,s\\
 &  & &  &  & 1\,$\times$\,350\,s\\
 &  & &  &  & 1\,$\times$\,440\,s\\
 \noalign{\smallskip}
\hline
\noalign{\smallskip}
GO-16748 & F. Niederhofer & 2021/09 & WFC3/UVIS & F814W & 2\,$\times$\,40\,s\\
&  &  &  & & 3\,$\times$\,446\,s\\
&  &  &  & & 2\,$\times$\,447\,s\\
\noalign{\smallskip}
\hline

\end{tabular}

\end{table*}


\begin{table*}
\centering
\caption{Observations of NGC 1856\label{tab:ngc1856obs}}
\begin{tabular}{@{}l@{ }c@{ }c@{ }c@{ }c@{ }c@{ }c@{ }}
\hline\hline
\noalign{\smallskip}
Programme ID & ~~~~~~~~PI~~~~~~~~ & ~~~~~~~Epoch~~~~~~~ & ~~~~~Instrument/Camera~~~~~ & ~~~~~Filter~~~~~ & ~~~~~Exposures~~~~~\\
& & (yyyy/mm) & & & (N\,$\times$\,t$_{\mathrm{exp}}$) \\
\noalign{\smallskip}
\hline
\noalign{\smallskip}
GO-13011 & T. Puzia & 2013/11 & WFC3/UVIS & F438W & 1\,$\times$\,185\,s \\
 &  &  &  &  & 2\,$\times$\,430\,s \\
 &  &  &  & F555W & 2\,$\times$\,350\,s \\
&  &  &  & F814W & 1\,$\times$\,51\,s \\
&  &  &  &  & 1\,$\times$\,360\,s \\
&  &  &  &  & 1\,$\times$\,450\,s \\
\hline
\noalign{\smallskip}
GO-13379 &  A. Milone & 2014/04 & WFC3/UVIS & F336W & 8\,$\times$\,711\,s \\
&   &  & & F814W & 4\,$\times$\,90\,s \\
&   &  & &  & 4\,$\times$\,704\,s \\
 \noalign{\smallskip}
\hline
\noalign{\smallskip}
GO-16748 & F. Niederhofer & 2022/07 & WFC3/UVIS & F814W & 2\,$\times$\,40\,s\\
&  &  &  & & 3\,$\times$\,446\,s\\
&  &  &  & & 2\,$\times$\,447\,s\\
\noalign{\smallskip}
\hline

\end{tabular}

\end{table*}


\begin{table*}
\centering
\caption{Observations of NGC 1866\label{tab:ngc1866obs}}
\begin{tabular}{@{}l@{ }c@{ }c@{ }c@{ }c@{ }c@{ }c@{ }}
\hline\hline
\noalign{\smallskip}
Programme ID & ~~~~~~~~PI~~~~~~~~ & ~~~~~~~Epoch~~~~~~~ & ~~~~~Instrument/Camera~~~~~ & ~~~~~Filter~~~~~ & ~~~~~Exposures~~~~~\\
& & (yyyy/mm) & & & (N\,$\times$\,t$_{\mathrm{exp}}$) \\
\noalign{\smallskip}
\hline
\noalign{\smallskip}
GO-14069 & N. Bastian & 2016/06 & WFC3/UVIS & F438W & 1\,$\times$\,95\,s \\
&  &  &  &  & 1\,$\times$\,150\,s \\
&  &  &  &  & 1\,$\times$\,450\,s \\
&  &  &  &  & 1\,$\times$\,500\,s \\
&  &  &  & F555W & 1\,$\times$\,75\,s \\
&  &  &  &  & 1\,$\times$\,95\,s \\
&  &  &  &  & 1\,$\times$\,150\,s \\
&  &  &  &  & 1\,$\times$\,450\,s \\
&  &  &  &  & 1\,$\times$\,500\,s \\
&  &  &  &  & 1\,$\times$\,550\,s \\
&  &  &  &  & 1\,$\times$\,700\,s \\
\hline
\noalign{\smallskip}
GO-14204 &  A. Milone & 2016/06 & WFC3/UVIS & F336W & 6\,$\times$\,711\,s \\
&   &  & & F814W & 3\,$\times$\,90\,s \\
&   &  & &  & 3\,$\times$\,678\,s \\
 \noalign{\smallskip}
\hline
\noalign{\smallskip}
GO-16748 & F. Niederhofer & 2022/09 & WFC3/UVIS & F814W & 2\,$\times$\,40\,s\\
&  &  &  & & 3\,$\times$\,446\,s\\
&  &  &  & & 2\,$\times$\,447\,s\\
\noalign{\smallskip}
\hline

\end{tabular}

\end{table*}


\begin{table*}
\centering
\caption{Observations of NGC 1868\label{tab:ngc1868obs}}
\begin{tabular}{@{}l@{ }c@{ }c@{ }c@{ }c@{ }c@{ }c@{ }}
\hline\hline
\noalign{\smallskip}
Programme ID & ~~~~~~~~PI~~~~~~~~ & ~~~~~~~Epoch~~~~~~~ & ~~~~~Instrument/Camera~~~~~ & ~~~~~Filter~~~~~ & ~~~~~Exposures~~~~~\\
& & (yyyy/mm) & & & (N\,$\times$\,t$_{\mathrm{exp}}$) \\
\noalign{\smallskip}
\hline
\noalign{\smallskip}
GO-14710 &  A. Milone & 2016/12 & WFC3/UVIS & F336W & 3\,$\times$\,831\,s \\
&   &  & & F814W & 1\,$\times$\,90\,s \\
&   &  & &  & 1\,$\times$\,666\,s \\
 \noalign{\smallskip}
\hline
\noalign{\smallskip}
GO-16748 & F. Niederhofer & 2021/09 & WFC3/UVIS & F814W & 2\,$\times$\,35\,s\\
&  &  &  & & 3\,$\times$\,433\,s\\
&  &  &  & & 2\,$\times$\,434\,s\\
\noalign{\smallskip}
\hline

\end{tabular}

\end{table*}


\begin{table*}
\centering
\caption{Observations of NGC 1898\label{tab:ngc1898obs}}
\begin{tabular}{@{}l@{ }c@{ }c@{ }c@{ }c@{ }c@{ }c@{ }}
\hline\hline
\noalign{\smallskip}
Programme ID & ~~~~~~~~PI~~~~~~~~ & ~~~~~~~Epoch~~~~~~~ & ~~~~~Instrument/Camera~~~~~ & ~~~~~Filter~~~~~ & ~~~~~Exposures~~~~~\\
& & (yyyy/mm) & & & (N\,$\times$\,t$_{\mathrm{exp}}$) \\
\noalign{\smallskip}
\hline
\noalign{\smallskip}
GO-13435 &  M. Monelli & 2014/01 & WFC3/UVIS & F336W & 2\,$\times$\,1035\,s \\
&   &  & & F438W & 2\,$\times$\,200\,s \\
&   &  & & F814W & 1\,$\times$\,100\,s \\
 \noalign{\smallskip}
\hline
\noalign{\smallskip}
GO-16748 & F. Niederhofer & 2022/02 & WFC3/UVIS & F814W & 1\,$\times$\,40\,s\\
&  &  &  & & 2\,$\times$\,447\,s\\
&  & 2022/09 & WFC3/UVIS & F814W & 2\,$\times$\,40\,s\\
&  &  &  &  & 2\,$\times$\,446\,s\\
&  &  &  &  & 2\,$\times$\,447\,s\\
\noalign{\smallskip}
\hline

\end{tabular}

\end{table*}


\begin{table*}
\centering
\caption{Observations of NGC 1978\label{tab:ngc1978obs}}
\begin{tabular}{@{}l@{ }c@{ }c@{ }c@{ }c@{ }c@{ }c@{ }}
\hline\hline
\noalign{\smallskip}
Programme ID & ~~~~~~~~PI~~~~~~~~ & ~~~~~~~Epoch~~~~~~~ & ~~~~~Instrument/Camera~~~~~ & ~~~~~Filter~~~~~ & ~~~~~Exposures~~~~~\\
& & (yyyy/mm) & & & (N\,$\times$\,t$_{\mathrm{exp}}$) \\
\noalign{\smallskip}
\hline
\noalign{\smallskip}
GO-9891 & G. Gilmore  & 2003/10 & ACS/WFC & F555W & 1\,$\times$\,300\,s \\
 &   &  &  & F814W & 1\,$\times$\,200\,s \\
\noalign{\smallskip}
\hline
\noalign{\smallskip}
GO-12257 & L. Girardi  & 2011/08 & WFC3/UVIS & F336W & 1\,$\times$\,660\,s \\
 &   &  &  & & 2\,$\times$\,740\,s \\
 &   & &  & F555W & 1\,$\times$\,60\,s \\
 &   &  &  & & 1\,$\times$\,300\,s \\
&   &  &  & & \,$\times$\,680\,s \\
\noalign{\smallskip}
\hline
\noalign{\smallskip}
GO-14069 & N. Bastian  & 2016/09 & WFC3/UVIS & F336W & 1\,$\times$\,380\,s \\
 &   &  & &  & 1\,$\times$\,460\,s \\
 &   &  & &  & 1\,$\times$\,740\,s \\
&   &  & & F438W & 1\,$\times$\,75\,s \\
&   &  & &  & 1\,$\times$\,120\,s \\
&   &  & &  & 1\,$\times$\,420\,s \\
&   &  & &  & 1\,$\times$\,460\,s \\
&   &  & &  & 1\,$\times$\,650\,s \\
&   &  & &  & 1\,$\times$\,750\,s \\
\noalign{\smallskip}
\hline
\noalign{\smallskip}
GO-15630 & N. Bastian  & 2019/09 & WFC3/UVIS & F814W & 3\,$\times$\,200\,s \\
 &   &  & &  & 3\,$\times$\,348\,s \\
  &   &  & &  & 1\,$\times$\,688\,s \\
\noalign{\smallskip}
\hline

\end{tabular}

\end{table*}


\begin{table*}
\centering
\caption{Observations of NGC 2005\label{tab:ngc2005obs}}
\begin{tabular}{@{}l@{ }c@{ }c@{ }c@{ }c@{ }c@{ }c@{ }}
\hline\hline
\noalign{\smallskip}
Programme ID & ~~~~~~~~PI~~~~~~~~ & ~~~~~~~Epoch~~~~~~~ & ~~~~~Instrument/Camera~~~~~ & ~~~~~Filter~~~~~ & ~~~~~Exposures~~~~~\\
& & (yyyy/mm) & & & (N\,$\times$\,t$_{\mathrm{exp}}$) \\
\noalign{\smallskip}
\hline
\noalign{\smallskip}
GO-12257 &  L. Girardi & 2011/10 & WFC3/UVIS & F475W & 1\,$\times$\,120\,s \\
&  &  & &  & 1\,$\times$\,600\,s \\
&  &  & &  & 1\,$\times$\,720\,s \\
&   &  & & F814W & 1\,$\times$\,30\,s \\
&   &  & & & 2\,$\times$\,700\,s \\
 \noalign{\smallskip}
\hline
\noalign{\smallskip}
GO-16748 & F. Niederhofer & 2022/07 & WFC3/UVIS & F814W & 2\,$\times$\,40\,s\\
&  &  &  &  & 3\,$\times$\,446\,s\\
&  &  &  &  & 2\,$\times$\,447\,s\\
\noalign{\smallskip}
\hline

\end{tabular}

\end{table*}


\begin{table*}
\centering
\caption{Observations of NGC 2108\label{tab:ngc2108obs}}
\begin{tabular}{@{}l@{ }c@{ }c@{ }c@{ }c@{ }c@{ }c@{ }}
\hline\hline
\noalign{\smallskip}
Programme ID & ~~~~~~~~PI~~~~~~~~ & ~~~~~~~Epoch~~~~~~~ & ~~~~~Instrument/Camera~~~~~ & ~~~~~Filter~~~~~ & ~~~~~Exposures~~~~~\\
& & (yyyy/mm) & & & (N\,$\times$\,t$_{\mathrm{exp}}$) \\
\noalign{\smallskip}
\hline
\noalign{\smallskip}
GO-9891 & G. Gilmore  & 2003/08 & ACS/WFC & F555W & 1\,$\times$\,250\,s \\
 &   &  &  & F814W & 1\,$\times$\,170\,s \\
 \noalign{\smallskip}
\hline
\noalign{\smallskip}
GO-10595 & P. Goudfrooij  & 2006/08 & ACS/WFC & F435W & 1\,$\times$\,90\,s \\
 &   &  &  &  & 2\,$\times$\,340\,s \\
 &   &  &  & F555W & 1\,$\times$\,40\,s \\
  &   &  &  &  & 2\,$\times$\,340\,s \\
 &   &  &  & F814W & 1\,$\times$\,8\,s \\
  &   &  &  &  & 2\,$\times$\,340\,s \\
 \noalign{\smallskip}
\hline
\noalign{\smallskip}
GO-16748 & F. Niederhofer & 2021/10 & WFC3/UVIS & F814W & 2\,$\times$\,40\,s\\
&  &  &  &  & 3\,$\times$\,446\,s\\
&  &  &  &  & 2\,$\times$\,447\,s\\
\noalign{\smallskip}
\hline

\end{tabular}

\end{table*}


\begin{table*}
\centering
\caption{Observations of NGC 2173\label{tab:ngc2173obs}}
\begin{tabular}{@{}l@{ }c@{ }c@{ }c@{ }c@{ }c@{ }c@{ }}
\hline\hline
\noalign{\smallskip}
Programme ID & ~~~~~~~~PI~~~~~~~~ & ~~~~~~~Epoch~~~~~~~ & ~~~~~Instrument/Camera~~~~~ & ~~~~~Filter~~~~~ & ~~~~~Exposures~~~~~\\
& & (yyyy/mm) & & & (N\,$\times$\,t$_{\mathrm{exp}}$) \\
\noalign{\smallskip}
\hline
\noalign{\smallskip}
GO-12257 &  L. Girardi & 2011/10 & WFC3/UVIS & F336W & 2\,$\times$\,700\,s \\
&  &  & &  & 1\,$\times$\,800\,s \\
&  &  & & F475W & 1\,$\times$\,120\,s \\
&  &  & & & 2\,$\times$\,700\,s \\
&   &  & & F814W & 1\,$\times$\,30\,s \\
&   &  & & & 1\,$\times$\,550\,s \\
&   &  & & & 2\,$\times$\,700\,s \\
 \noalign{\smallskip}
\hline
\noalign{\smallskip}
GO-16748 & F. Niederhofer & 2022/06 & WFC3/UVIS & F814W & 2\,$\times$\,40\,s\\
&  &  &  &  & 4\,$\times$\,454\,s\\
&  &  &  &  & 1\,$\times$\,455\,s\\
\noalign{\smallskip}
\hline

\end{tabular}

\end{table*}


\begin{table*}
\centering
\caption{Observations of NGC 2203\label{tab:ngc2203obs}}
\begin{tabular}{@{}l@{ }c@{ }c@{ }c@{ }c@{ }c@{ }c@{ }}
\hline\hline
\noalign{\smallskip}
Programme ID & ~~~~~~~~PI~~~~~~~~ & ~~~~~~~Epoch~~~~~~~ & ~~~~~Instrument/Camera~~~~~ & ~~~~~Filter~~~~~ & ~~~~~Exposures~~~~~\\
& & (yyyy/mm) & & & (N\,$\times$\,t$_{\mathrm{exp}}$) \\
\noalign{\smallskip}
\hline
\noalign{\smallskip}
GO-12257 &  L. Girardi & 2011/10 & WFC3/UVIS & F336W & 2\,$\times$\,700\,s \\
&  &  & &  & 1\,$\times$\,800\,s \\
&  &  & & F475W & 1\,$\times$\,120\,s \\
&  &  & & & 2\,$\times$\,700\,s \\
&   &  & & F814W & 1\,$\times$\,30\,s \\
&   &  & & & 1\,$\times$\,550\,s \\
&   &  & & & 2\,$\times$\,700\,s \\
 \noalign{\smallskip}
\hline
\noalign{\smallskip}
GO-16748 & F. Niederhofer & 2022/12 & WFC3/UVIS & F814W & 2\,$\times$\,40\,s\\
&  &  &  &  & 5\,$\times$\,453\,s\\
\noalign{\smallskip}
\hline

\end{tabular}

\end{table*}


\begin{table*}
\centering
\caption{Observations of NGC 2209\label{tab:ngc2209obs}}
\begin{tabular}{@{}l@{ }c@{ }c@{ }c@{ }c@{ }c@{ }c@{ }}
\hline\hline
\noalign{\smallskip}
Programme ID & ~~~~~~~~PI~~~~~~~~ & ~~~~~~~Epoch~~~~~~~ & ~~~~~Instrument/Camera~~~~~ & ~~~~~Filter~~~~~ & ~~~~~Exposures~~~~~\\
& & (yyyy/mm) & & & (N\,$\times$\,t$_{\mathrm{exp}}$) \\
\noalign{\smallskip}
\hline
\noalign{\smallskip}
GO-12908 &  P. Goudfrooij & 2013/04 & WFC3/UVIS & F438W & 2\,$\times$\,850\,s \\
&   &  & & F814W & 1\,$\times$\,60\,s \\
&   &  & & & 2\,$\times$\,485\,s \\
 \noalign{\smallskip}
\hline
\noalign{\smallskip}
GO-16748 & F. Niederhofer & 2022/07 & WFC3/UVIS & F814W & 2\,$\times$\,40\,s\\
&  &  &  &  & 4\,$\times$\,454\,s\\
&  &  &  &  & 1\,$\times$\,455\,s\\
\noalign{\smallskip}
\hline

\end{tabular}

\end{table*}


\begin{table*}
\centering
\caption{Observations of NGC 2210\label{tab:ngc2210obs}}
\begin{tabular}{@{}l@{ }c@{ }c@{ }c@{ }c@{ }c@{ }c@{ }}
\hline\hline
\noalign{\smallskip}
Programme ID & ~~~~~~~~PI~~~~~~~~ & ~~~~~~~Epoch~~~~~~~ & ~~~~~Instrument/Camera~~~~~ & ~~~~~Filter~~~~~ & ~~~~~Exposures~~~~~\\
& & (yyyy/mm) & & & (N\,$\times$\,t$_{\mathrm{exp}}$) \\
\noalign{\smallskip}
\hline
\noalign{\smallskip}
GO-14164 &  A. Sarajedini & 2017/04 & WFC3/UVIS & F336W & 3\,$\times$\,700\,s \\
&   &  & &  & 4\,$\times$\,715\,s \\
&   &  & &  & 4\,$\times$\,729\,s \\
&   &  & &  & 4\,$\times$\,730\,s \\
&  & 2016/11 & ACS/WFC & F606W & 2\,$\times$\,50\,s \\
&  &  &  & & 6\,$\times$\,348\,s \\
&  &  &  & & 6\,$\times$\,353\,s \\
&  & 2016/11 & ACS/WFC & F814W & 2\,$\times$\,70\,s \\
&  &  &  &  & 6\,$\times$\,344\,s \\
&  &  &  &  & 6\,$\times$\,378\,s \\
&  &  &  &  & 6\,$\times$\,413\,s \\
 \noalign{\smallskip}
\hline
\noalign{\smallskip}
GO-16748 & F. Niederhofer & 2022/12 & WFC3/UVIS & F814W & 2\,$\times$\,40\,s\\
&  &  &  & & 3\,$\times$\,446\,s\\
&  &  &  & & 2\,$\times$\,447\,s\\
\noalign{\smallskip}
\hline

\end{tabular}

\end{table*}


\begin{table*}
\centering
\caption{Observations of NGC 2213\label{tab:ngc2213obs}}
\begin{tabular}{@{}l@{ }c@{ }c@{ }c@{ }c@{ }c@{ }c@{ }}
\hline\hline
\noalign{\smallskip}
Programme ID & ~~~~~~~~PI~~~~~~~~ & ~~~~~~~Epoch~~~~~~~ & ~~~~~Instrument/Camera~~~~~ & ~~~~~Filter~~~~~ & ~~~~~Exposures~~~~~\\
& & (yyyy/mm) & & & (N\,$\times$\,t$_{\mathrm{exp}}$) \\
\noalign{\smallskip}
\hline
\noalign{\smallskip}
GO-12257 &  L. Girardi & 2011/11 & WFC3/UVIS & F475W & 1\,$\times$\,120\,s \\
&   &  & & & 1\,$\times$\,600\,s \\
&   &  & & & 1\,$\times$\,720\,s \\
&   &  & & F814W & 1\,$\times$\,30\,s \\
&   &  & & & 2\,$\times$\,700\,s \\
 \noalign{\smallskip}
\hline
\noalign{\smallskip}
GO-16748 & F. Niederhofer & 2022/03 & WFC3/UVIS & F814W & 2\,$\times$\,40\,s\\
&  &  &  &  & 4\,$\times$\,454\,s\\
&  &  &  &  & 1\,$\times$\,455\,s\\
\noalign{\smallskip}
\hline

\end{tabular}

\end{table*}


\begin{table*}
\centering
\caption{Observations of NGC 2257\label{tab:ngc2257obs}}
\begin{tabular}{@{}l@{ }c@{ }c@{ }c@{ }c@{ }c@{ }c@{ }}
\hline\hline
\noalign{\smallskip}
Programme ID & ~~~~~~~~PI~~~~~~~~ & ~~~~~~~Epoch~~~~~~~ & ~~~~~Instrument/Camera~~~~~ & ~~~~~Filter~~~~~ & ~~~~~Exposures~~~~~\\
& & (yyyy/mm) & & & (N\,$\times$\,t$_{\mathrm{exp}}$) \\
\noalign{\smallskip}
\hline
\noalign{\smallskip}
GO-14164 &  A. Sarajedini & 2016/12 & WFC3/UVIS & F336W & 8\,$\times$\,700\,s \\
&   &  & &  & 4\,$\times$\,705\,s \\
&   &  & &  & 3\,$\times$\,733\,s \\
&  & 2016/02 & ACS/WFC & F606W & 2\,$\times$\,50\,s \\
&  &  &  & & 3\,$\times$\,364\,s \\
&  &  &  & & 6\,$\times$\,353\,s \\
&  &  &  & & 2\,$\times$\,525\,s \\
&  & 2016/02 & ACS/WFC & F814W & 2\,$\times$\,70\,s \\
&  &  &  &  & 6\,$\times$\,363\,s \\
&  &  &  &  & 3\,$\times$\,390\,s \\
&  &  &  &  & 6\,$\times$\,400\,s \\
&  &  &  &  & 2\,$\times$\,450\,s \\
 \noalign{\smallskip}
\hline
\noalign{\smallskip}
GO-16748 & F. Niederhofer & 2022/03 & WFC3/UVIS & F814W & 2\,$\times$\,35\,s\\
&  &  &  & & 3\,$\times$\,433\,s\\
&  &  &  & & 3\,$\times$\,434\,s\\
\noalign{\smallskip}
\hline

\end{tabular}

\end{table*}

\end{appendix}

\end{document}